\documentclass[aps,12pt,prd,superscriptaddress,preprintnumbers, 
	amssymb,
	amsmath,
	notitlepage,
	longbibliography,
	nofootinbib]{revtex4-1}
\usepackage[colorlinks=true, 
	linkcolor=blue, 
	citecolor=blue, 
	urlcolor=blue, 
	linktocpage=true]{hyperref}
\usepackage[utf8]{inputenc}
\usepackage[english]{babel}
\usepackage{lmodern}
\usepackage{graphicx,color,float}
\usepackage{wrapfig}

\usepackage{mathtools}

\usepackage[
]{todonotes}

\usepackage[framemethod=TikZ]{mdframed}
\usepackage{lipsum}

\mdfdefinestyle{MyFrame}{
linecolor=blue,
outerlinewidth=2pt,
roundcorner=20pt,
innertopmargin=\baselineskip,
innerbottommargin=\baselineskip,
innerrightmargin=20pt,
innerleftmargin=20pt,
backgroundcolor=gray!20!white}

\newcommand{\be}{\begin{equation}}
\newcommand{\ee}{\end{equation}}
\newcommand{\beal}{\begin{aligned}}
\newcommand{\eeal}{\end{aligned}}
\newcommand\bea {\begin{eqnarray}}
\newcommand\eea {\end{eqnarray}}
\newcommand{\bec}{\begin{cases}}
\newcommand{\eec}{\end{cases}}
\newcommand{\bei}{\begin{itemize}}
\newcommand{\eei}{\end{itemize}}
\newcommand{\bee}{\begin{enumerate}}
\newcommand{\eee}{\end{enumerate}}

\definecolor{remFcol}{RGB}{0,140,220}

\begin{document}

\title{Higgs Vacuum Decay from Particle Collisions?}
\date{\today}

\author{Leopoldo Cuspinera}
\email{j.l.cuspinera@durham.ac.uk}
\affiliation{Centre for Particle Theory, Durham University,
South Road, Durham, DH1 3LE, UK}
\author{Ruth Gregory}
\email{r.a.w.gregory@durham.ac.uk}
\affiliation{Centre for Particle Theory, Durham University,
South Road, Durham, DH1 3LE, UK}
\affiliation{Perimeter Institute, 31 Caroline Street North, Waterloo, 
ON, N2L 2Y5, Canada}
\author{Katie M. Marshall}
\email{k.marshall6@newcastle.ac.uk}
\affiliation{School of Mathematics, Statistics and Physics, Newcastle University, 
Newcastle Upon Tyne, NE1 7RU, UK}
\author{Ian G. Moss}
\email{ian.moss@newcastle.ac.uk}
\affiliation{School of Mathematics, Statistics and Physics, Newcastle University, 
Newcastle Upon Tyne, NE1 7RU, UK}

\begin{abstract}
We examine the effect of large extra dimensions on black hole seeded
vacuum decay using the Randall-Sundrum model as a prototype for warped
extra dimensions. We model the braneworld black hole by a tidal solution,
and solve the Higgs equations of motion for the instanton on the brane. 
Remarkably, the action of the static instanton can be shown to be the
difference in the bulk areas of the seed and remnant black holes, and we
estimate these areas assuming the black holes are small compared to the
bulk AdS radius. Comparing to the Hawking evaporation rate shows that 
small black hole seeds preferentially catalyse vacuum decay, thus extending
our previous results to higher dimensional braneworld scenarios. The
parameter ranges do not allow for Standard Model Higgs decay from
collider black holes, but they can be relevant for cosmic ray collisions.
\end{abstract}

\preprint{DCPT-18/05}

\maketitle

\section{Introduction}

A fascinating consequence of the discovery of the Higgs 
\cite{ATLAS:2012ae,Chatrchyan:2012tx}, is that the standard model 
vacuum appears to be metastable \cite{Degrassi:2012ry,
Gorsky:2014una,Bezrukov:2014ina,Ellis:2015dha,Blum:2015rpa}
(see also earlier work \cite{Krive:1976sg,1982Natur.298,
Sher:1988mj,Isidori:2001bm,EliasMiro:2011aa}). Although it was originally thought 
that this would not be an issue due to the extremely long half-life 
predicted by the classic bubble nucleation arguments of 
Coleman et al.\ \cite{coleman1977,callan1977,CDL}, (see also
\cite{Kobzarev:1974cp}), recent work by two of us
\cite{GMW,BGM1,BGM2,BGM3,Gregory:2016xix} indicates that
the situation may not be quite so rosy.
In \cite{GMW}, we developed
a description of vacuum decay catalysed by black holes,
with the result that the strong local spacetime curvature of
small black holes catalyses vacuum decay and
dramatically changes the prediction for the lifetime of the 
universe\footnote{Some of these results were examined in \cite{Tetradis:2016vqb}, 
however without explicitly computing the Euclidean instanton action.}.
Tunnelling is initiated by a black hole seed in the the false vacuum 
that decays into a remnant black hole surrounded by Higgs fields 
which have overcome the potential barrier and lie in a lower energy
state. The tunnelling rate is determined by the difference in action between
the remnant black hole-instanton combination and the seed black hole
false vacuum configuration that turns out to be proportional to the difference
in horizon area of the seed and remnant black holes. Because of
this dependence on black hole area, enhancement occurs
only for very small black holes, the obvious candidates being
primordial black holes in our universe, indeed, there is an
interesting thermal interpretation of our result, see for example
\cite{Chen:2017suz,Gorbunov:2017fhq,Mukaida:2017bgd}.

There is however another possible scenario in which small black
holes could occur, and that is in particle collisions. If we have a situation
where our four dimensional Planck scale is derived from a higher 
dimensional Planck mass close to the standard model scale
\cite{ArkaniHamed:1998rs,Antoniadis:1998ig,Randall:1999ee,
Randall:1999vf}, then
it is easier to form black holes in particle collisions 
\cite{Giddings:2001bu,Dimopoulos:2001hw,Landsberg:2003br,Harris:2004xt}. 
Such higher dimensional 
theories are dubbed \emph{Large Extra Dimension} scenarios, 
and the premise is that we live
on a four dimensional ``brane'' in a higher dimensional spacetime.
Our relatively high Planck scale, $M_p = 1/\sqrt{8\pi G_N}$, is the result of a 
geometric hierarchy coming from an integration over the extra dimensions.
Since the true Planck scale is the higher dimensional one, 
it is easier to form black holes in high energy processes,
leading to the possibility of black holes being
produced at the LHC (for a review see \cite{Park:2012fe}).
Given this exciting possibility for producing small black holes, 
we should revisit our four dimensional black hole instanton 
calculations and explore the impact of large extra dimensions.

As a first step in looking at vacuum decay with extra dimensions,
we considered the impact of dimensionality on our toy model thin 
wall calculations in \cite{BGM2}, finding that extra dimensions seemed 
to impede vacuum decay, however, these estimates were predicated 
on a rather crude higher dimensional generalization that did not take 
the braneworld aspect of the Large Extra Dimension models into account. 
In this paper, we revisit the role of large
extra dimensions in vacuum decay, explicitly modelling the brane black 
hole and finding exact solutions for the instanton on the brane.
We also make a more careful estimation of the black hole
Hawking radiation rate on the brane. We find that, while for a
given seed mass the higher dimensional tunnelling rate is
indeed lower than the four dimensional one, what we gain from
higher dimensions is that lower seed masses are allowed due to the
lower value of the fundamental Planck scale, $M_D$.

The layout of the paper is as follows: in the next section, we review
the status of constructing instantons both in four dimensions
with black holes, and for braneworlds in five dimensions without
black holes, and discuss the problems involved in introducing
a black hole to the higher dimensional calculation. In section
\ref{sec:action} we discuss the calculation of the action of an
approximate black hole instanton, showing that, as in four
dimensions, the static instanton action is the difference in black hole
horizon areas. In section \ref{sec:bubble} we solve for the brane
scalar field and find the instantons and their actions numerically.
In section \ref{sec:disc} we conclude.

\section{Braneworlds and Black Holes}

It is perhaps worth recalling the various challenges in finding 
an instanton for vacuum decay in a braneworld setting.
The braneworld paradigm describes our universe as an effective
submanifold of a higher dimensional manifold, with standard
model fields living only on the four-dimensional braneworld, but
with gravity propagating throughout all of the dimensions, leading
to the renormalization of Newton's constant. For one extra
dimension we can consistently solve for the spacetime geometry
using the Israel approach \cite{Israel:1966}, giving the standard
Randall-Sundrum (RS) braneworld \cite{Randall:1999vf}, a 
paradigm for warped compactifications.
For higher codimension, there is no unique ``delta-function'' limit for
a thin braneworld \cite{Geroch:1987qn}, and typically one resorts
to approximate hybrid Kaluza-Klein/warped descriptions for gravity 
on a lower-dimensional brane. Thus, for a concrete gravitational 
description in this paper we will remain within the RS model.

The RS model supposes that we have one extra 
dimension, and that the higher dimensional spacetime, or bulk,
has a negative cosmological constant. The braneworld has
a positive tension, and the vacuum brane has an energy-momentum
tensor that is parallel to the brane with energy and tension equal.
The original solution presented by Randall and Sundrum had the tension
tuned to give a flat brane:
\be
ds^2 = e^{-2|z|/\ell} \eta_{\mu\nu} dx^\mu dx^\nu - dz^2
\ee
where the cusp in the warp factor at $z=0$ corresponds to the 
brane. The local negative curvature of the bulk supports the brane 
tension that is easily calculated from the Israel junction conditions:
\be
{\cal K}^{(+)}_{\mu\nu} = -\frac1\ell \eta_{\mu\nu} \;\; \Rightarrow\;\;\;
8\pi G_5 \sigma = \Delta {\cal K}_{\mu\nu} -  \Delta {\cal K} \eta_{\mu\nu}
= \frac6\ell \eta_{\mu\nu}
\ee
and is tuned to fit with the cosmological constant $\Lambda_5 = -6/\ell^2$. 
De-tuned branes, with tension greater or less than this critical
value may also be embedded within the bulk AdS spacetime,
although the natural embeddings now become either space- or time-like
\cite{Chamblin:1999ya,Kaloper:1999sm,Kraus:1999it,Binetruy:1999ut,
BCG,Karch:2000gx}, but as long as the brane energy-momentum 
is approximately homogeneous (i.e.\ having a spatially isotropic
pressure term only) the bulk solution can be fully integrated, and
the brane trajectory found \cite{BCG}.

For a brane black hole solution, we must break this spatial homogeneity,
but even with the added benefit of having only one codimension, the
exact solution for a brane black hole has been extremely elusive
\cite{Gregory:2008rf,Kanti:2009sz}.
The natural geometry of a Schwarzschild black hole that
extends off the brane into a black string, found by Chamblin,
Hawking and Reall \cite{Chamblin:1999by},
has the problem that it is neither representative of matter
localised on the brane, nor is it stable, suffering from a Gregory-Laflamme type 
of instability \cite{Gregory:1993vy,Gregory:2000gf}. A lower dimensional
analogue of the brane black hole was found by Emparan et al.\
\cite{Emparan:1999wa,Emparan:1999fd} by taking a $2+1$ dimensional
brane through the equatorial plane of a $4$-dimensional AdS C-metric
\cite{Kinnersley:1970zw,Plebanski:1976gy}.
The black hole would be expected to be accelerating from the perspective
of the bulk, since an observer hovering at fixed distance from the brane
is in fact undergoing uniform acceleration towards it. Unfortunately,
there is no known exact solution for a C-metric in more than 4 dimensions,
thus no template for constructing a braneworld black hole plus bulk analytically.

To maintain an analytic approach one can explore the 
effective brane gravitational equations using the approach of 
Shiromizu et al.\ \cite{Shiromizu:1999wj}, leading to the tidal solution 
that we will use in this paper \cite{Dadhich:2000am}. (One can also
explore braneworlds with additional matter, either on the brane or in the
bulk, to support analyticity of the brane embedding, see e.g.\ 
\cite{Galfard:2005va,Creek:2006je,Dai:2010jx,Kanti:2013lca})
Alternately, one can take a numerical approach; the equations of motion 
to be solved are an elliptic system \cite{Wiseman:2001xt}, with
the brane junction conditions and asymptotic Poincare
horizon providing the boundary conditions. The solutions for small
black holes were found  in \cite{Kudoh:2003xz}, although the
large black hole solutions have been far more tricky to 
determine due to the nonlinearity of the Einstein equations and
the impact of the bulk warping of the horizon, however there has
been some interesting recent work in this direction 
\cite{Figueras:2011gd,Wang:2016nqi}.

\begin{wrapfigure}[23]{r}{0.45\textwidth}
\centering
\vskip -5mm
\includegraphics[width=0.3\textwidth]{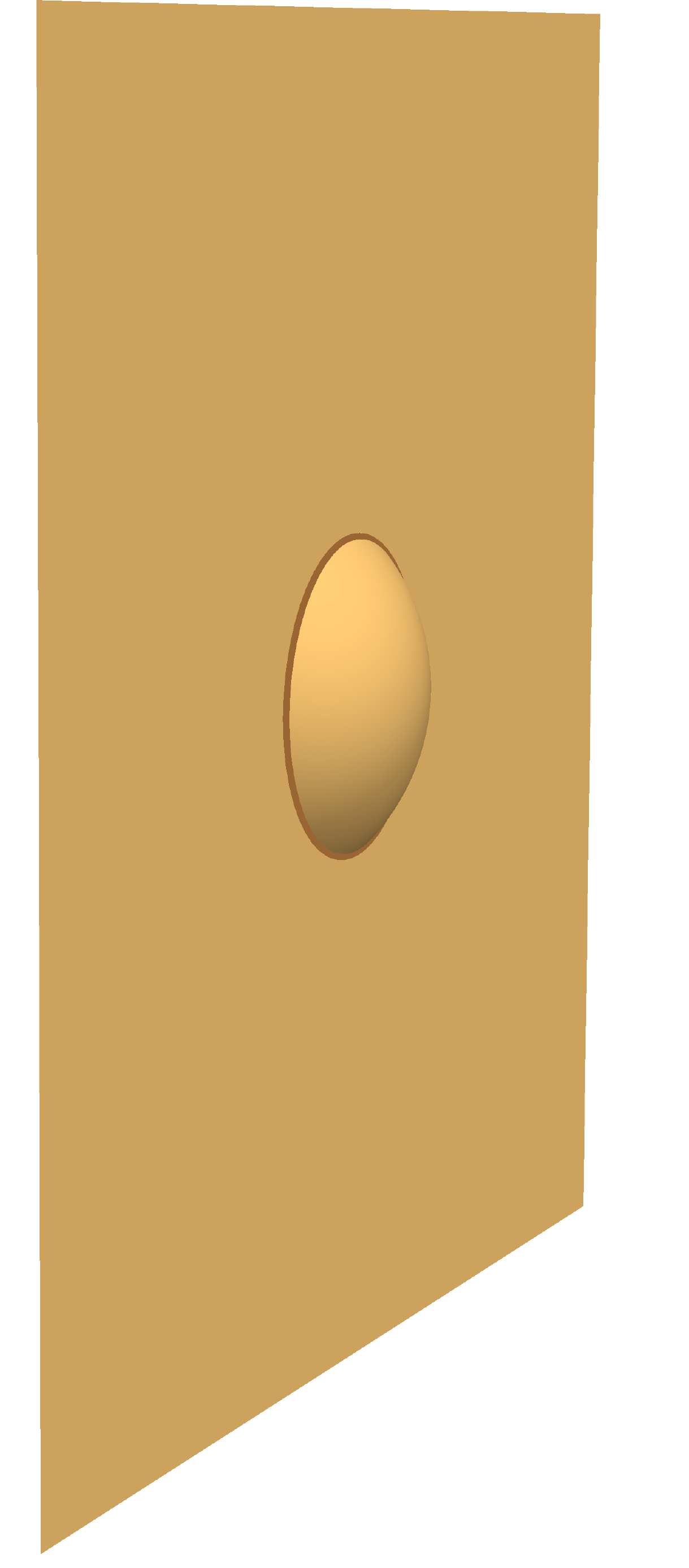}
\caption{
The braneworld instanton for decay of a Minkowski false vacuum 
brane to a sub-critical AdS brane from \cite{Gregory:2001dn}. }
\label{fig:simplecdl}
\end{wrapfigure}
Now let us consider the instanton from a higher dimensional perspective. 
The decay of a metastable false vacuum was first computed by Coleman
and collaborators in a series of papers \cite{coleman1977,callan1977,CDL}, 
in which a Euclidean approach was used to find an instanton solution
interpolating between the true and false vacua. A convenient approximation,
extremely useful for visualisation, is to take the region over which the 
vacuum interpolates to be very narrow in comparison with the interior
of the bubble. This ``thin wall'' then has a straightforward generalisation
to gravity, as described in the paper with de Luccia \cite{CDL} (CDL).
While this thin wall description is not appropriate for the Higgs
vacuum decay \cite{BGM3}, where the vacuum interpolation is 
very wide and relatively gentle, it nonetheless provides an excellent
shorthand for visualising the process of decay. 

The CDL picture however, is very symmetric, and assumes that both the 
initial and final states are completely devoid of features and are 
homogeneous. If instead one relaxes this assumption, minimally, by
allowing for an inhomogeneity in the form of a black hole, the analytic
approach of CDL can be preserved, and the equations of motion for
the instanton are only minimally altered \cite{GMW,BGM1,BGM2,BGM3},
however, the impact on the action of the instanton can be quite
significant, and particularly for the thick scalar domain walls
appropriate to the Higgs potential \cite{BGM3}, tunnelling
turns out to be significantly enhanced to the extent that if there
are primordial black holes, false vacuum decay will happen.

Let us now consider how these arguments might lift to higher 
dimensions. In \cite{Gregory:2001dn}, the equivalent of the CDL
instantons on a Randall-Sundrum braneworld were constructed,
the 5D instanton being geometrically akin to the 4D representations
of the CDL instantons. Sub- and super- critical branes follow
spherical trajectories in the AdS bulk, so the tunneling of a Minkowski
false vacuum to an AdS true vacuum is represented by a flat
brane with a bubble sticking out, as shown in figure \ref{fig:simplecdl}.
As is usual with the RS model, two copies of the picture are identified,
and the ``bubble wall'' is the sharp edge between the spherical and 
flat parts of the braneworld,
appearing roughly as a codimension two object. 

\begin{wrapfigure}[21]{r}{0.45\textwidth}
\centering
\vskip -5mm
\includegraphics[width=0.25\textwidth]{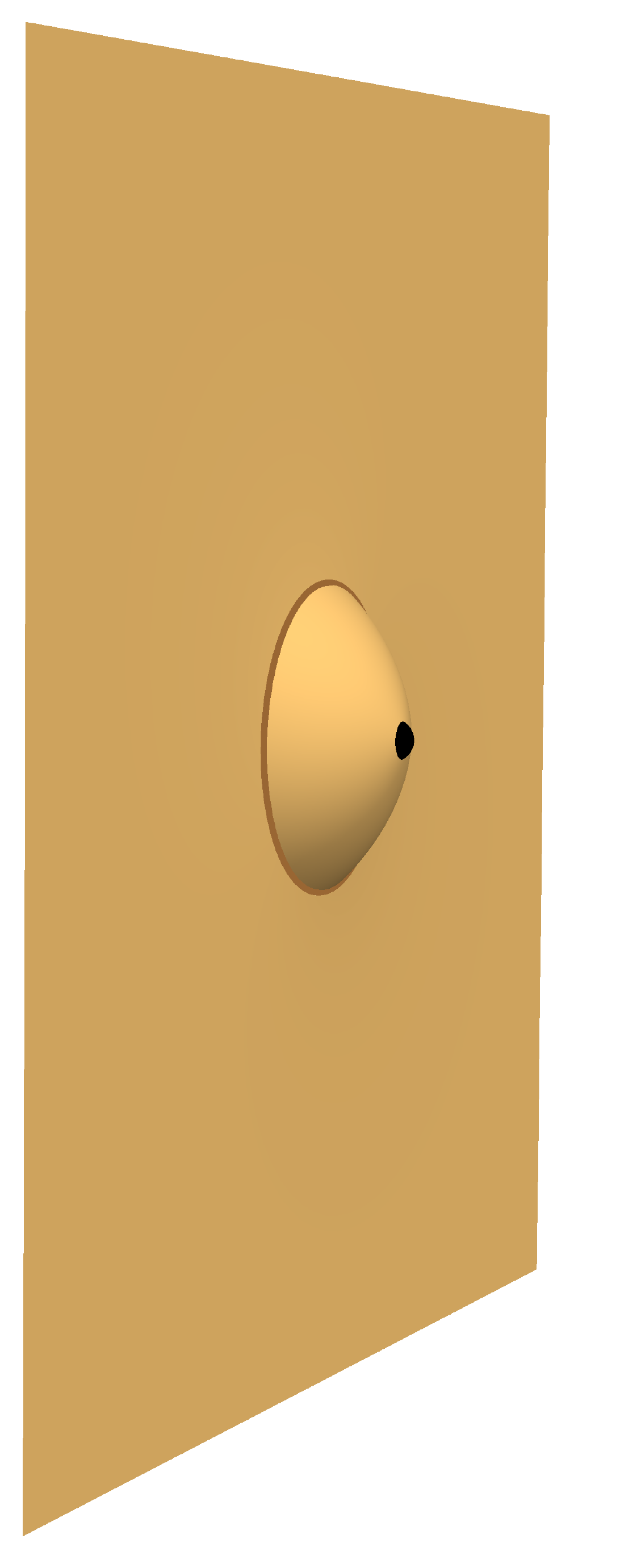}
\caption{A sketch of the expected geometry of braneworld
vacuum decay with a braneworld black hole.}
\label{fig:bhcdl}
\end{wrapfigure}
Ideally, one would like to construct a similar instanton, but with a 
black hole, however, at this point the lack of an exact brane black
hole solution becomes problematic. Even if we drop a dimension
to have a 2+1 dimensional braneworld, for which the brane black
hole solution is constructed via the C-metric \cite{Emparan:1999wa},
we have the problem that the C-metric has a unique slicing for the 
braneworld \cite{Kudoh:2004ub}, so we cannot patch together two 
different braneworld trajectories such as an equatorial sub-critical
slice matching to a flat brane further away as suggested in figure \ref{fig:bhcdl}. 
Indeed, slicing a bulk Schwarzschild metric induces additional energy 
momentum on the brane \cite{Galfard:2005va,Creek:2006je},
(except for the uniform radius ``cosmological'' brane solutions).

Thus as a direct approach to finding the instanton seems problematic, 
we follow a more pragmatic approach, and rather than seeking an exact
analytic solution, instead consider what a black hole instanton might 
approximately look like. From the intuition gleaned in the 4D black hole
instantons, we expect that small black holes are the most dangerous,
and that the dominant instanton will be the static instanton \cite{BGM3}.
Then, analogous to the modelling of collider black hole phenomenology
\cite{Harris:2003db}, we
use the higher dimensional Schwarzschild-AdS solution as an
approximation to the local bulk black hole: this allows us to 
construct a method of calculating the instanton action formally.
Finally, in order to correctly identify the asymptotics of our instanton,
we need a way of interpolating between the near horizon 
and far-field brane solution, which we expect to have a 4D
Schwarzschild $G_NM/r$ behaviour. This final step requires 
a choice for the braneworld solution, and we use the 
tidal brane solution of Dadhich et al.\ \cite{Dadhich:2000am},
found by considering vacuum solutions with a non-vanishing
bulk Weyl tensor in the formalism of Shiromizu et al.\ \cite{Shiromizu:1999wj}.
The tidal solution has the attractive feature that it has 
the correct asymptotic form at large brane radius, but looks like
the five dimensional Schwarzschild potential for small radius,
indeed, it is similar to the Reissner-Nordstrom black hole,
although the ``tidal charge'' term $-r_Q^2/r^2$ is negative. This 
tidal charge was not related to the mass in \cite{Dadhich:2000am},
but left as an arbitrary degree of freedom, therefore part of our task 
in section \ref{sec:bubble} will be to relate the tidal charge to the
mass of the black hole. 

Our strategy is then as follows: we first take our brane black hole,
approximately modelled by the 5D Sch-AdS solution, and continue
to Euclidean time. We then compute the action of this solution 
in a rather general way, using the approach of Hawking and Horowitz
\cite{Hawking:1995fd};
as per usual, the direct way of computing the action leads to an
apparent divergence that we cannot in this case regulate directly
by introducing a cut-off as we will explain. Nonetheless, however 
we choose to regulate the action, the same method will apply for
the false vacuum black hole and the instanton bubble 
solution, thus we simply subtract the seed and bubble actions
to get the final amplitude for vacuum decay. Crucially, this turns
out to be simply the difference in areas of the seed and remnant
black hole horizon geometries. Finally, we integrate the scalar 
equations of motion on the brane to obtain the brane bubble solution,
and use the tidal metric to relate the near horizon and asymptotic
geometries. The nett result is an amplitude for brane black hole seeded
vacuum decay that we can compare to the higher dimensional
brane black hole evaporation rate to explore whether brane vacuum
metastability is an issue.

\section{The Euclidean Brane Black Hole Action}
\label{sec:action}

In this section we will show that, just like in four dimensions, the 
Euclidean action of any static black hole solution can be expressed 
entirely by surface terms. This is a remarkable result, because it not 
only applies to the vacuum black hole, it also applies with a cosmological 
constant, with matter and even with a conical singularity at the horizon.

We begin by recalling the properties of the 
Euclidean Schwarzschild black hole in four dimensions
\begin{equation}
ds^2=f(r)d\tau^2+f(r)^{-1}dr^2+r^2d\Omega_{I\!I}^2,
\end{equation}
where 
\begin{equation}
f(r)=1-\frac{2G_NM}{r}
\end{equation}
In order to explore the geometry near the `horizon' $r_h=2G_NM$, 
we expand using a new coordinate ${\varrho}$, defined by
\begin{equation}
{\varrho}=\sqrt{{2(r-r_h)\over\kappa}}
\label{varrho4d}
\end{equation} 
where $\kappa$ is the surface gravity, $\kappa=f'(r_h)/2$.
To leading order $f(r)= \kappa^2{\varrho}^2+O({\varrho}^4)$, and
close to the horizon,
\begin{equation}
ds^2=d{\varrho}^2+{\varrho}^2d(\kappa\tau)^2
+r_h^2d\Omega_{I\!I}^2+O({\varrho}^4),
\end{equation}
For small ${\varrho}\ge 0$, the metric is geometrically the product of a 
disc with a sphere, provided that $\kappa\tau$ is taken to be an angular 
coordinate with the usual range $2\pi$. If $\kappa\tau$
has a different range, then the manifold has a conical singularity at 
$r_h$. Note that the Euclidean section is perfectly regular other than
this, but only covers the exterior region of the original black hole. 
The \emph{event} horizon of the original Lorentzian black hole is 
encoded in the topology of the Euclidean solution: the surface 
${\varrho}=0$ is a 2-sphere of radius $r_h$.

For the brane black hole in five dimensions, the metric is extended
into an additional direction, parametrised by $\chi$ in Kudoh et al.\
\cite{Kudoh:2003xz}, who numerically constructed small brane black
holes with horizon size less than the AdS radius $\ell$. In
\cite{Kudoh:2003xz}, the metric was written in the form
\be
ds^2 = \frac{1}{(1+\frac{\rho}{\ell}\cos\chi)^2} \left [ T^2(\rho,\chi) d\tau^2 
+ e^{2B(\rho,\chi)} (d\rho^2+\rho^2d\chi^2)
+ e^{2C(\rho,\chi)} \rho^2 \sin^2\!\chi d\Omega_{I\!I}^2 \right]\,,
\label{fivemetric}
\ee
where the brane sits at $\chi=\pi/2$, and $\chi\leq\pi/2$ is kept as the bulk.
Clearly, in the small black hole limit, $\ell \to \infty$, we have the five 
dimensional Schwarzschild black hole:
\be
ds^2 = \left (\frac{\rho^2-\rho_h^2}{\rho^2+\rho_h^2} \right)^2 d\tau^2
+ \left (\frac{\rho^2+\rho_h^2}{\rho^2} \right)^2 \left [
d\rho^2 + \rho^2 d\Omega_{I\!I\!I}^2 \right]
\ee
written here in homogeneous co-ordinates, rather than the area
gauge. The local Euclidean horizon coordinate is 
${\varrho}=2(\rho-\rho_h)$, and the horizon has area
${\cal A} = 4\rho_h^2$, and surface gravity
\be
\kappa = e^{-B(\rho_h)} T'
\label{Kudohkappa}
\ee
The black hole is corrected at order $\rho/\ell$ by the conformal factor, 
and at order $\rho_h/\ell$ in the other metric functions close to the horizon. 
Kudoh and collaborators integrated the functions $T, B$ and $C$ 
numerically, and found that the $T$ function to a very good approximation
extends hyperspherically off the brane. Although $B$ and $C$ are not
precisely the same, their difference is roughly of order $\rho_h/\ell$
as expected. At large $\rho$, $T,B,C\to1$, and 
the metric is asymptotically AdS in the Poincar\'e patch.

We do not use the explicit form of the metric, however, the features
we require from the solutions of \cite{Kudoh:2003xz} are that the
event horizon is topologically hyperspherical with
constant surface gravity, and that the
braneworld black hole asymptotes the Poincar\'e patch of AdS.
The coordinate transformation between the local black hole 
coordinates and the Poincar\'e RS coordinates is
\be
\rho^2 = r^2 + \ell^2 (e^{|z|/\ell}-1)^2,\qquad
\tan\chi = \frac{r}{\ell(e^{|z|/\ell}-1)}\,,
\ee
and we expect that the `trajectory' of the brane in the
black hole metric will bend slightly in response to the black
hole at $\rho_h$, giving rise to a four dimensional Newtonian
potential as described in \cite{Garriga:1999yh}. From the
perspective of the $\{\rho,\chi\}$ coordinates, in which the 
brane sits at $\chi=\pi/2$, this will show up as a $1/\rho$ correction
to $T,B,C$. We therefore take our asymptotic metric to be of the form
\be
ds^2 =e^{-2|z|/\ell}\left [ F(r,z)d\tau^2
+ F(r,z)^{-1}dr^2+r^2d\Omega^2\right]+dz^2,
\label{asympmetric}
\ee
where $F\sim1-2G_NM(z)/r+O(r^{-2})$. We can think of $M(z)$ as
coming from the brane bending term of $M/\rho$ in the original coordinates.

\subsection{Computing the Action}

\begin{figure}[htb]
\begin{center}
\includegraphics[width=0.4\textwidth]{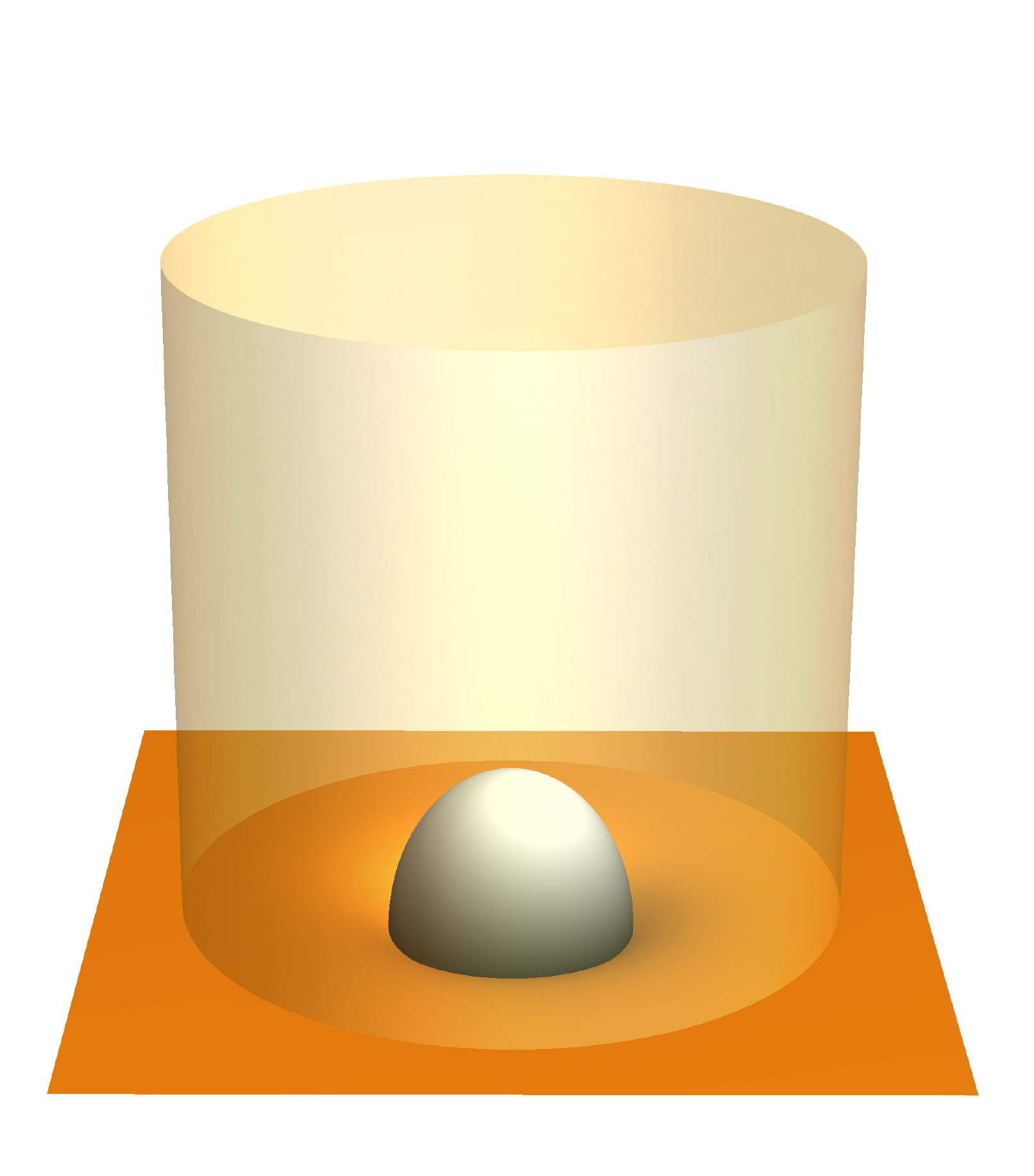}~~
\includegraphics[width=0.4\textwidth]{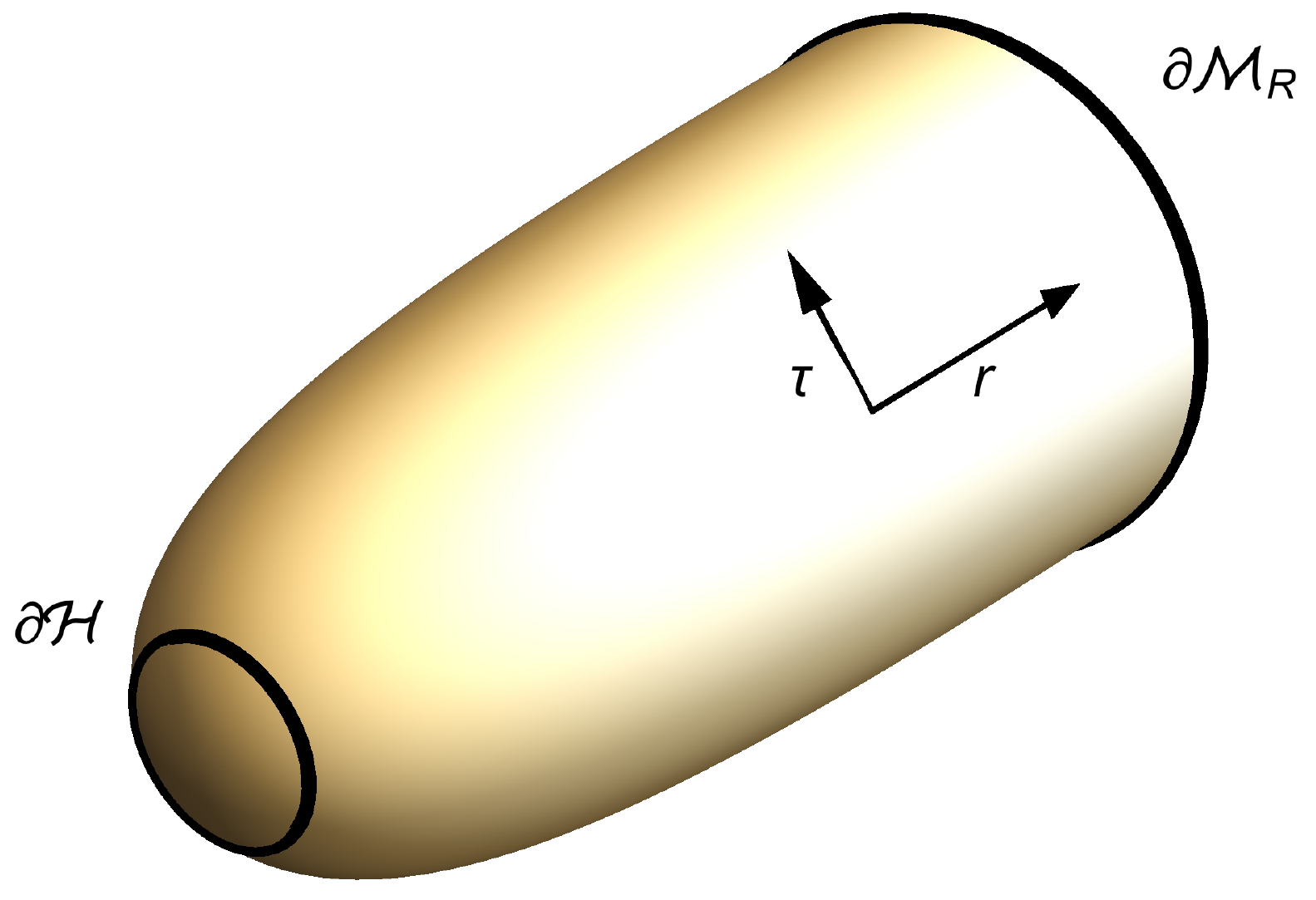}
\caption{A cartoon of the Euclidean tidal black hole and the cut-off surfaces.
On the left, the $\tau, \theta$ coordinates are suppressed, and the cut-off
surface is indicated relative to the brane and bulk black hole horizon. 
Only one half of the $\mathbb{Z}_2$ symmetric solution is shown.
On the right, the Euclidean $\tau$ coordinate is shown but the bulk and angular
coordinates are suppressed, and the ``black hole cigar'' geometry is indicated. 
Two circles denote the boundary ${\partial \cal H}$ of the region just outside the 
horizon and the boundary ${\partial \cal M}_r$ at large radius.}
\label{fig:cartoon}
\end{center}
\end{figure}

The action of the black hole instanton combination diverges and has to 
be regulated in some way. We do this by truncating the five dimensional
manifold at large distances from the black hole, taking a surface at
large radius $R$ on the brane, and extending this along geodesics 
in the $\pm z$ directions orthogonal to the brane to produce the outer 
boundary surface ${\partial\cal M}_R$ as indicated in the cartoon in 
figure \ref{fig:cartoon}. The interior is denoted by ${\cal M}_R$ 
and the intersection of ${\cal M}_R$ with the brane world is 
denoted by ${\cal B}$.

The Euclidean action for this truncated instanton or black hole
solution is
\begin{equation}
I_R=-\frac{1}{16\pi G_5}\int_{{\cal M}_R}(R_5-2\Lambda_5)
+\int_{{\cal B}}{\cal L}_m\sqrt{g}
+\frac{1}{8\pi G_5}\int_{{\partial\cal M}_R} K\sqrt{h},
\end{equation}
where $K$ denotes the extrinsic curvature of the boundary surface
$\partial {\cal M}_R$ defined with an \emph{inward} pointing normal
to the bulk manifold ${\cal M}_R$. The matter Lagrangian ${\cal L}_m$ includes
the contribution from any nontrivial Higgs field profile, as well as
the brane stress-energy tensor. The bulk integral is understood to 
range across all $z$, and includes the $\delta-$function curvature
at the brane source in the spirit of the Israel approach.
Note that the gravitational constant in five dimensions is related to 
Newton's constant in four dimensions by $G_5=\ell G_N$.

We now show that the tunnelling exponent, given by the difference
between the actions of the instanton geometry with a remnant black
hole, and the false vacuum geometry with the seed black hole:
$B=I_{\text{inst}}-I_{\text{FV}}$, is finite in the limit $R\to\infty$. 
The first step is to introduce a small ball, ${\cal H}$, extending a 
proper distance of order ${\cal O}(\varepsilon)$ out from the black 
hole event horizon, to formally deal with any conical deficits arising 
from a generic periodicity in Euclidean time. This splits the action 
calculation into two terms,
\begin{equation}
I_R=I_{R}^{\rm hor}+I_{R}^{\rm ext},
\end{equation}
where\footnote{Note, the extrinsic curvature in the Gibbons-Hawking term 
is computed with an inward pointing normal, hence the \emph{same} sign
for that term in each expression.}
\be
I_R^{\rm hor}=-\frac{1}{16\pi G_5}\int_{\cal H}(R_5-2\Lambda_5)+
\int_{{\cal B}_{\cal H}}{\cal L}_m\sqrt{g} +
\frac{1}{8\pi G_5}\int_{\partial\cal H} K\sqrt{h},\hskip 15mm
\ee
\be
\beal
I_R^{\rm ext}&=-\frac{1}{16\pi G_5}\int_{{\cal M}_R-{\cal H}}(R_5-2\Lambda_5)+
\int_{{\cal B}-{\cal B}_{\cal H}}{\cal L}_m\sqrt{g}
+\frac{1}{8\pi G_5}\int_{\partial\cal H} K\sqrt{h}\\
&\qquad+\frac{1}{8\pi G_5}\int_{{\partial\cal M}_R} K\sqrt{h},
\eeal
\label{ir}
\ee
and ${\cal B}_{\cal H}= {\cal B}\cap{\cal H}$ is the intersection of the 
event horizon cap with the brane.

In order to deal with the near-horizon contribution, we transform 
\eqref{fivemetric} to local horizon coordinates,
analogous to the Euclidean Schwarzschild transformation, 
\eqref{varrho4d}, so that
\begin{equation}
ds^2\approx d\varrho^2+A^2(\varrho,\xi)d\tau^2
+D^2(\varrho,\xi) d\Omega^2_{I\!I}+N^2(\varrho,\xi)d\xi^2,
\label{nearhor}
\end{equation}
where $\varrho<\varepsilon$ inside ${\cal H}$.
Comparing to \eqref{fivemetric}, we see $A = T/(1+\frac\rho\ell\cos\chi)$,
$D = \rho \sin\chi e^C/(1+\frac\rho\ell\cos\chi)$, with $\varrho \approx 
(\rho-\rho_h)/(1+\frac{\rho_h}\ell\cos\chi)$
and $\xi = \chi + {\cal O}(\varrho^2)$. The brane sits at $\xi=\pi/2$,
and on the horizon, $\xi\in[0,\pi]$.

As with the four dimensional Euclidean Schwarzschild, there is
a natural periodicity of $\tau$ for which the Euclidean metric
is nonsingular; this periodicity is $\beta_0=2\pi/\kappa$, where
$\kappa$ is the surface gravity of the black hole given in the original
coordinates by \eqref{Kudohkappa}, and in the horizon coordinates 
by $\partial A/\partial\varrho$. From nonsingularity of the geometry, we deduce
$N\sim N_0(\xi) + {\cal O}(\varrho^2)$, $D\sim D_0(\xi)+{\cal O}(\varrho^2)$, 
and $A \sim \kappa \varrho+ {\cal O}(\varrho^2)$. Now let us consider a
general periodicity $\beta$ for the Euclidean time $\tau$, then we will have 
a conical singularity at $\varrho=0$. In order to compute the action, we smooth
this out by modifying the $A$ function so that $A'(\varepsilon,\xi) = \kappa$,
but $A'(0,\xi) = \kappa \beta_0/\beta$. Computing the curvature for this
smoothed metric gives
\be
\sqrt{g} (R-2\Lambda_5) = - 2 N_0(\xi) C_0(\xi)^2 A'' (\varrho) + {\cal O}(\varrho)
\ee
which gives the bulk contribution to $I_R^{\rm hor}$ as
\be
\beal
-\frac{1}{16\pi G_5}\int_{\cal H}(R_5-2\Lambda_5)+
\int_{{\cal B}_{\cal H}}{\cal L}_m\sqrt{g}&=
\frac{\beta}{2} [A'(\varepsilon)-A'(0)] \int N_0D_0^2 d\xi 
+ {\cal O}(\varepsilon^2)\\
&= \frac{\kappa}{8\pi} [\beta-\beta_0] {\cal A}_5
\eeal
\ee
where $ {\cal A}_5=4\pi \int N_0D_0^2 d\xi $ is the area of the braneworld
black hole horizon extending into the bulk (on both sides of the brane).

To compute the Gibbons-Hawking boundary term we note that the normal to
${\partial \cal H}$ is $n = -d\varrho$, hence the extrinsic curvature is
\be
K=-A^{-1}A_{,\varrho}+{\cal O}(\varepsilon) 
\ee
and
\be
\frac{1}{8\pi G_5} \int_{\partial\cal H} K\sqrt{h}
= -\frac{\kappa \beta}{2G_5} \int N_0D_0^2 d\xi 
= -\frac{\kappa \beta {\cal A}_5}{8\pi G_5}
\ee
Thus the contribution to the action from the horizon region is
\be
I_R^{\rm hor}=-\frac{\kappa\beta_0 {\cal A}_5}{8\pi G_5}
= -\frac{{\cal A}_5}{4G_5}
\ee

In appendix \ref{Appaction}, we show that the external part 
$I_R^{\rm ext}$ can be simplified by taking a canonical decomposition 
based on a foliation of the manifold by surfaces of constant $\tau$, 
$\Sigma_\tau$, and the part of the action outside the horizon cylinder 
reduces to simple surface terms,
\be
I_R^{\rm ext} = \frac{1}{8\pi G_5} \int_0^\beta d\tau 
\left( \int_{C_R}  {}^3\!K\, \sqrt{h} + \int_{C_{\cal H}}  {}^3\!K\, \sqrt{h} \right).
\label{baction}
\ee
where ${}^3\!K$ are the extrinsic curvatures of codimension two
surfaces of constant $r$, regarded as submanifolds of 
surfaces of constant $\tau$, $\Sigma_\tau$ as described in appendix
\ref{Appaction}.

Close to the horizon, we use the metric \eqref{nearhor} and find
\begin{equation}
{}^3\!K=2D^{-1}D_{,\varrho}+N^{-1}N_{,\varrho}\to 0,
\end{equation}
at the horizon $\varrho=0$ for the behaviour of the metric coefficients 
$D(\varrho,\xi)$ and $N(\varrho,\xi)$ given earlier. There is no contribution
to the action from this boundary term.

At large distances, the metric approaches the perturbed Poincar\'e 
form \eqref{asympmetric}, and we find
\be
{}^3\!K=-\frac{2}{R}e^{|z|/\ell} F^{1/2},\qquad
\sqrt{h}=R^2e^{-3|z|/\ell} F^{1/2}.
\ee
hence
\be
I_R^{\rm ext}=-\frac{\beta}{G_N\ell}\int_0^\infty \! dz
e^{-2z/\ell}\left(2R-4G_NM(z)+O(R^{-1})\right).
\ee
Ideally, we would like to regularise this action either by background
subtraction, or adding in boundary counterterms along the lines of
\cite{Balasubramanian:1999re,Emparan:1999pm}, however, the 
counterterms of \cite{Emparan:1999pm} do not regulate this action, 
and one cannot replace the interior of ${\cal M}_R$ with a pure RS
braneworld, due to the variation of $M(z)$ along $\partial {\cal M}_R$.
Instead, we note that the Higgs fields on the brane in any instanton
solution will die off exponentially for large $r$, so from the intuition 
that $M(z)/r \sim M_\infty/\rho = M_\infty/\sqrt{r^2 + \ell^2 (e^{|z|/\ell}-1)^2}$, 
we then deduce that the mass function $M(z)$ will be the same at leading 
order for both the false vacuum with the seed brane black hole, and the 
instanton solution, therefore the exterior terms will cancel when we take the
difference between the instanton action and the false vacuum action:
\be
B = I_{\text{inst}} - I_{\text{FV}} = \lim_{R\to\infty} \left [
I_R^{\rm ext} \Big|_{\text{inst}} -I_R^{\rm ext} \Big|_{\text{FV}} \right]
-\frac{{\cal A}_5^{\text{inst}}}{4G_5}
+\frac{{\cal A}_5^{\text{FV}}}{4G_5}
= \frac{{\cal A}_5^{\rm seed}}{4G_5}
-\frac{{\cal A}_5^{\rm rem}}{4G_5}
\label{bterm}
\ee
where ${\cal A}_5^{\rm seed}$ and ${\cal A}_5^{\rm rem}$ refer to the areas of the seed 
and remnant black hole horizon areas respectively.

This is simply the reduction in entropy $-\Delta S$ caused by the decay 
process, and the tunnelling rate is recognisable as the probability of an 
entropy reduction $\propto \exp(\Delta S)$.
The difficulty we face when applying (\ref{bterm}) is that we have to relate the
black hole area to the mass of the black hole triggering the vacuum decay and the
physical parameters in the Higgs potential. This requires explicit solutions for the
gravitational and Higgs fields.

\section{Tidal black hole bubbles}
\label{sec:bubble}

As we reviewed, the main obstacle to finding tunnelling instantons is the 
lack of any analytic brane black hole solutions. The brane-vacuum
equations are complicated by the reduced symmetry of the expected
static, brane-rotationally symmetric geometry. Although we
have numerical brane black hole solutions, once we introduce Higgs
profiles on the brane, these would be modified, and a new full
numerical brane$+$bulk solution would have to be computed --
a formidable task. Instead, we adopt a more practical alternative, 
based on the tidal black hole solutions of Dadhich et al.\ \cite{Dadhich:2000am}.

As described, for example, by Maartens \cite{Maartens:2000fg}, one can take
an approach of solving purely the brane ``Einstein equations'',
i.e.\ the induced Einstein equations on the brane found by the Gauss
Codazzi projection of the Einstein tensor in Shiromizu et al.\  
\cite{Shiromizu:1999wj} (SMS). These equations are similar to the four
dimensional Einstein equations, but contain additional terms involving
the square of the energy momentum of any matter on the brane, and
an additional so-called Weyl tensor, ${\cal E}_{\mu\nu}$, 
coming from a projection of the bulk Weyl tensor onto the brane.
The Weyl tensor for the tidal black hole satisfies the equations 
${\cal E}_\mu{}^\mu=0$ and $\nabla^\mu{\cal E}_{\mu\nu}=0$. 
Following \cite{Maartens:2000fg}, one uses the symmetry of the
physical set up to write the Weyl tensor as
\be
{\cal E}^\mu_\nu = \text{diag} \left ( {\cal U} , -\frac{({\cal U} + 2\Pi)}{3},
\frac{\Pi - {\cal U}}{3} \right)
\ee
This is manifestly tracefree, and the `Bianchi' identity 
implies a conservation equation for ${\cal U}, \Pi$. For the spherically
symmetric static brane metric
\be
ds^2_{\text{brane}} = f(r) e^{2\delta(r)}d\tau^2
+f^{-1}(r) dr^2 + r^2 d\Omega^2_{I\!I},
\label{genbranemet}
\ee
the conservation equation implies
\be
\left ( {\cal U} + 2\Pi \right) ' + \left ( \frac{f'}{f} + 2\delta' \right)
\left ( 2{\cal U} + \Pi \right) +  \frac{6\Pi }{r} = 0\,.
\label{weyleos}
\ee
Even for the vacuum brane this is not a closed system, but if
one assumes an equation of state, one can find an induced 
brane solution \cite{Gregory:2004vt}. The tidal black hole 
corresponds to the choice $\Pi = -2{\cal U}$, for which
\eqref{weyleos} is easily solved by ${\cal U} \propto 1/r^4$.

The tidal black hole of Dadhich et al.\ \cite{Dadhich:2000am}, 
has $\delta(r) \equiv 0$, 
\be
f(r)=1-\frac{2G_NM}{r}-\frac{r_Q^2}{r^2}\,,
\label{vbh}
\ee
and
\be
{\cal E}_{\mu\nu}dx^\mu dx^\nu=-\frac{r_Q^2}{ r^4}\left(
f(r) d\tau^2+f^{-1}(r) dr^2-r^2d\Omega^2\right),
\label{conf}
\ee
where $r_Q$ is a constant parameter related to the tidal charge $Q$
of \cite{Dadhich:2000am} by $r_Q^2 = -Q$. The motivation
for this solution is clear: at large distances, the Newtonian potential
of a mass source has the conventional $G_NM/r$ behaviour due to 
a ``brane-bending'' term identified by Garriga and Tanaka \cite{Garriga:1999yh}; 
the interpretation being that the brane shifts relative to the bulk in 
response to matter on the brane. At small distances on the other hand 
we would expect the higher dimensional Schwarzschild potential to be 
more appropriate, hence the $-r_Q^2/r^2$ term. The event horizon is 
distorted by the Weyl tensor, hence the name. 
Other choices for the Weyl tensor
lead to different brane solutions \cite{Gregory:2004vt}, however these 
tend to have either wormholes or singularities (or both), therefore we
do not consider these here.

For our bubble solution, we will need to find the fully coupled 
Higgs plus brane SMS-gravitational equations of motion in the 
spherically symmetric gauge \eqref{genbranemet}, and we
will use the same \emph{tidal} Ansatz for the equation of state
of the Weyl tensor: $\Pi = -2{\cal U}$. The beauty of the tidal Ansatz
is that even with the Higgs fields taking a nontrivial bubble profile, 
the conservation equation for the Weyl tensor \eqref{weyleos}
is still solved by ${\cal U} = - r_Q^2/r^4$.

We also have some limited information about the form of the tidal 
black hole solution away from the brane from an expansion in the 
fifth coordinate. According to Maartens and Koyama, \cite{Maartens:2010ar}
the metric parallel to the brane at proper distance $z$ from the brane is
\be
\tilde g_{\mu\nu}(z) = g_{\mu\nu}(0) - ( 8\pi G_5 S_{\mu\nu}) \, z
+\left [ (4\pi G_5)^2 S_{\mu\sigma}S^\sigma{}_\nu
- 8\pi G_N S_{\mu\nu} - {\cal E}_{\mu\nu} \right] z^2 +\dots
\ee
where $S_{\mu\nu}=T_{\mu\nu}-\frac13 T g_{\mu\nu}$ is composed of 
the energy momentum tensor of brane matter.
In the false vacuum state, we have $T_{\mu\nu}=0$ and the metric 
expansion away from the brane reduces to
\be
\beal
ds^2&\approx e^{-2|z|/\ell}\left(g_{\mu\nu}-{\cal E}_{\mu\nu}z^2\right)+dz^2\\
& \approx e^{-2|z|/\ell}\left\{
\left(1 + \frac{r_Q^2z^2}{r^4}\right) \left( fd\tau^2+f^{-1}dr^2\right)
+ \left(1-\frac{r_Q^2z^2}{r^4}\right) r^2d\Omega^2_{I\!I} \right\}+dz^2,
\eeal
\ee
which shows clearly how the horizon area decreases in the 
$z$ direction. The horizon forms into a true bulk black hole when 
the area vanishes for some value of $z$ of order $r_h^2/r_Q$.

Although this tidal black hole has many attractive features, the
main difficulty that has to be overcome when finding the bubble 
solutions is that the tidal constant $r_Q$ is undetermined. 
Clearly a nonsingular brane black hole, if approximately tidal,
should have a relation between the asymptotic mass 
measured on the brane, $M$, and the tidal charge $r_Q^2$.
For very large black holes, we expect the horizon radius to be 
predominantly determined by $M$, and this ambiguity is not
relevant, however for the small black holes we are interested in,
the horizon radius is primarily dependent on $r_Q$, and we
must confront this ambiguity.

We start by noting that the tidal black hole solution should be 
identical to the five dimensional Schwarzschild black hole 
in the limit that the AdS radius $\ell\to\infty$, as the brane
stress-energy tensor, which is tuned to the cosmological constant, 
vanishes in this limit, and full $SO(4)$ rotational symmetry is restored. 
Since $G_N=G_5/\ell$, \eqref{vbh} implies that $r_Q^2\to r_h^2$
in this limit. Intuitively, we also expect that for small black holes, the 
bulk AdS scale should also be subdominant, and the black hole should
look (near the horizon at least) mainly like a five dimensional black
hole, i.e.\ $r_Q^2\to r_h^2$ as $r_h\to 0$. We will therefore assume 
analyticity in $r_h/\ell$ and write
\be
r_Q^2=r_h^2\left(1-b{r_h\over\ell}
+ {\cal O} \left({r_h^2\over\ell^2}\right)\right)\label{rq}
\ee
for small $r_h/\ell$, where $b$ is some constant independent of $r_h$
and $\ell$, expected to be roughly of order unity. For the tidal black hole, 
a trivial rewriting of \eqref{vbh} gives the relation
\be
M = \frac{b r_h^2}{2G_5}
\label{mtorh}
\ee
in other words, we have expressed the ambiguity in the tidal parameter 
for small black holes by the parameter $b$, and the relationship 
between the asymptotic mass of the black hole as measured on the brane 
and the horizon radius explicitly factors in this ambiguity. As we now see, 
this uncertainty can be absorbed into a redefinition of the low energy 
Planck scale in the tunnelling rate.

The tunnelling process starts with the uniform false vacuum $\phi_v$ 
and a seed black hole with mass $M_S$. This false vacuum configuration
resembles the tidal black hole on the brane, and a slightly perturbed
5D Schwarzschild solution in the bulk \cite{Kudoh:2003xz}. 
The bubble solution represents the decay process to another state 
with the field asymptoting the same false vacuum at large distances
but with the field approaching its true vacuum near the horizon of a
remnant black hole with mass $M_R$, which remains after tunnelling.

In the previous section we showed that the tunnelling exponent is given by
\begin{equation}
B=\frac{1}{4G_5}\left({\cal A}_S-{\cal A}_R\right),
\end{equation}
where $S$ represents the seed black hole area
and $R$ that of the remnant black hole (recall, this
area is the full five dimensional area of the horizon
extending into the bulk).
To leading order in $r_h/\ell$, the small black hole horizon has 
an approximately hyperspherical shape, therefore the area will
be well approximated by $2\pi^2 r^3$, hence
\begin{equation}
B= \frac{\pi^2}{2G_5}\left(r_S^3-r_R^3\right)
=  \frac{\pi^2 r_S^3}{2G_5}
\left[1- \left (\frac{M_R}{M_S} \right)^{\frac32}\right]
\end{equation}
using \eqref{mtorh}. In the limit that the difference in seed and
remnant black hole masses is small, $(M_S-M_R)/M_S = 
\delta M/M_s \ll 1$, we finally arrive at
\be
B\approx \frac34 \left(\frac{\pi M_S}{ b M_5}\right)^{3/2}
\frac{\delta M}{M_S},
\label{Bapprox}
\end{equation}
where  $M_5=(8\pi G_N\ell)^{-1/2}$ is the low energy Planck scale. Fortuitously,
the uncertainty in the value of the tidal charge parameter 
$b$ can be absorbed into our uncertainty in the low energy Planck scale, and 
so we let $bM_5\to M_5$.

\subsection{Higgs bubbles on the brane}

The Higgs bubble will correspond to a solution of the brane SMS
equations with an energy momentum tensor derived from the 
(Euclidean) scalar field Lagrangian\footnote{Note that we have
defined the Euclidean Lagrangian to contain $+V$, meaning that
the false vacuum solution will have energy-momentum $-V g_{\mu\nu}$,
but that our 4D Einstein equations will have the conventional sign
for the energy-momentum, i.e.\ $G_{\mu\nu} = 
8\pi G_N T_{\mu\nu}+ \dots$.}
\be
{\cal L}_m=\frac12g^{\mu\nu}\phi_{,\mu}\phi_{,\nu}+V(\phi).
\ee
where $V(\phi)$ has a metastable false vacuum. The SMS equations 
for the bubble, assuming the general form \eqref{genbranemet}
are derived in appendix \ref{app:braneq}, and are
\begin{align}
&f\phi''+f'\phi'+\frac2r f\phi'+f\delta'\phi'-V_{,\phi}=0,\label{eq1}\\
&\mu'=4\pi r^2\left\{ \frac12 f\phi'{}^2+V-
\frac{2\pi G_N}{3}\ell^2(\frac12 f\phi'{}^2-V)
(\frac32 f\phi'{}^2+V)\right\},\label{eq2}\\
&\delta'=4\pi G_N r\phi'{}^2\left\{1-\frac{4\pi G_N}{3}\ell^2
(\frac12 f\phi'{}^2-V)\right\}\label{eq3}.
\end{align}
where, for comparison with the vacuum case (\ref{vbh}), we 
have defined a ``mass'' function $\mu(r)$ by
\be
f(r)=1-\frac{2G_N\mu(r)}{r} - \frac{r_Q^2}{r^2}.
\label{vbub}
\ee
These are integrated numerically from the black hole horizon $r_h$ 
to $r\to\infty$ where $\phi$ is in the false vacuum.

\begin{figure}[htb]
\centering
\includegraphics[width=0.7\textwidth]{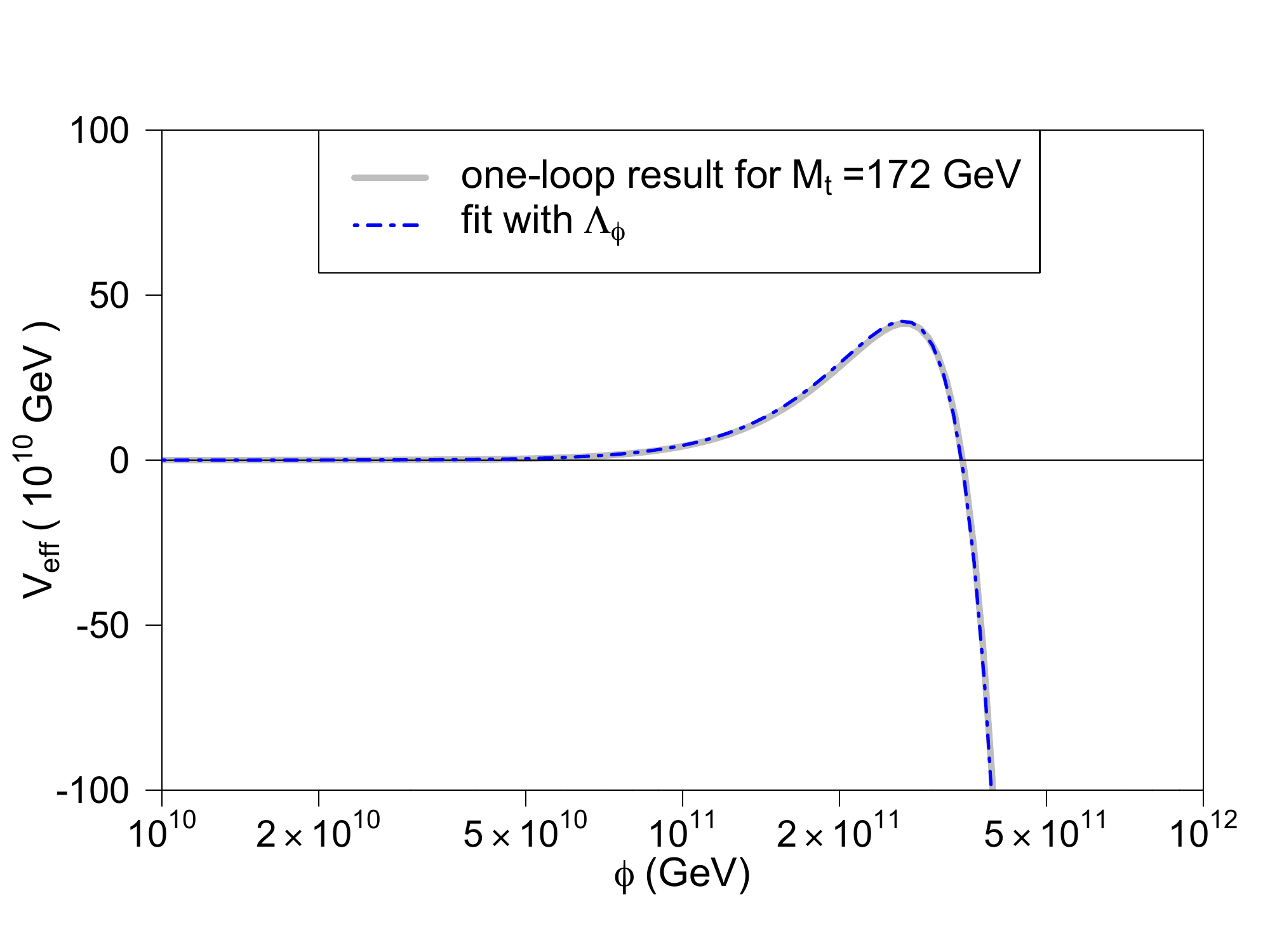}
\caption{The Higgs potential calculated numerically at one loop order for top
quark mass $M_t=172\,{\rm GeV}$ and the
approximate potential using (\ref{lambdaeff}) with values of $g$ and 
$\Lambda_\phi$ chosen
for the best fit.}
\label{fig:potential}
\end{figure}

The numerical results contained in this section are based on a Higgs-like potential,
assuming that the standard model holds for energy scales up to the low energy
Planck mass $M_5$. The detailed form of the potential is determined by
renormalisation group methods and depends on low-energy particle masses,
with strong dependence on the Higgs and top quark masses. Of these, the
top quark mass is less well known, and for masses in the range
$171-174\, {\rm GeV}$, Higgs instability sets in at scales from 
$10^{10}-10^{18}\,{\rm GeV}$.

The Higgs potential is usually expressed in the form 
\begin{equation}
V(\phi)=\frac14\lambda_{\rm eff}(\phi)\phi^4.
\end{equation}
with a running coupling constant $\lambda_{\rm eff}(\phi)$ that
becomes negative at some crossover scale $\Lambda_\phi$.
Vacuum decay depends on the shape of the potential barrier
in the Higgs potential around this instability scale, and in order 
to explore the likelihood of decay it is useful to use an analytic
fit to $\lambda_{\rm eff}$. In \cite{BGM3}, we used a two
parameter fit to $\lambda_{\rm eff}$, where one of the parameters
was closely related to the crossover scale. We found that the dependence
of the instanton action on the potential was strongly dependent on 
this parameter, but very weakly dependent on the second parameter,
which was more related to the shape of the potential at low energy. 
For clarity therefore, here we take a one parameter analytic fit to
$\lambda_{\rm eff}$, where the single parameter is the crossover scale
$\Lambda_\phi$:
\be
\lambda_{\rm eff}=g(\Lambda_\phi) \left\{\left(\ln \frac{\phi}{M_p}\right)^4
-\left(\ln \frac{\Lambda_\phi}{M_p}\right)^4\right\}
\label{lambdaeff}
\end{equation}
and $g(\Lambda_\phi)$, chosen to fit the high energy asymptote of 
$\lambda_{\rm eff}$, varies very little across the range of $\Lambda_\phi$
of relevance to the Standard Model $\lambda_{\rm eff}$. Figure
\ref{fig:potential} shows a sample of our analytic fit for the Higgs
potential to the actual $\lambda_{\rm eff}$ computed for $M_t = 172$GeV.
In four dimensions, we can have a Higgs instability scale very close to
the Planck scale, however with large extra dimensions, new physics 
could potentially enter at the low-energy Planck scale $M_5$, thus to be
consistent, we should restrict our parameters to the range $\Lambda_\phi<M_5<M_p$.
\begin{figure}[htb]
\centering
\includegraphics[width=0.9\textwidth]{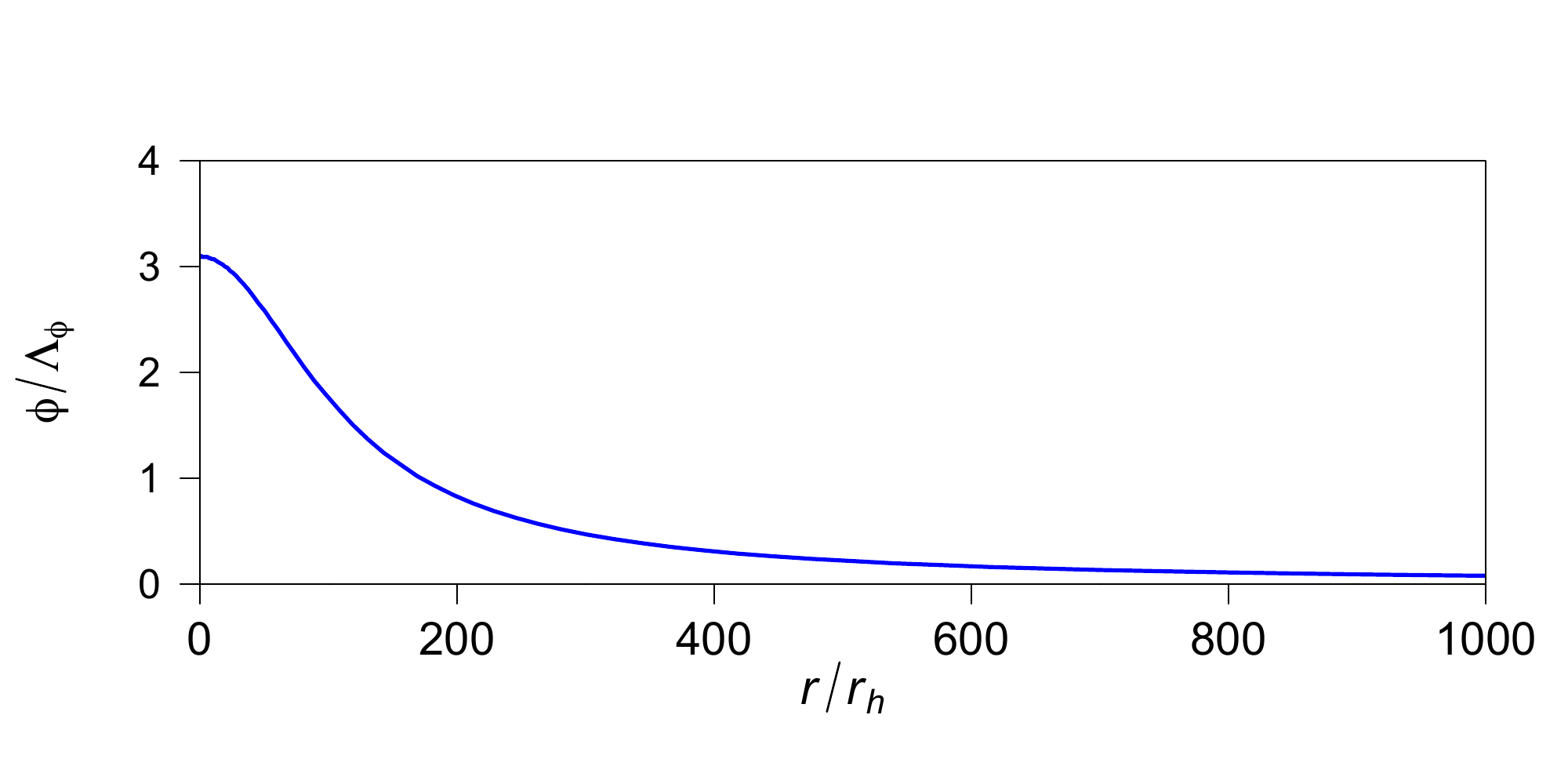}
\includegraphics[width=0.9\textwidth]{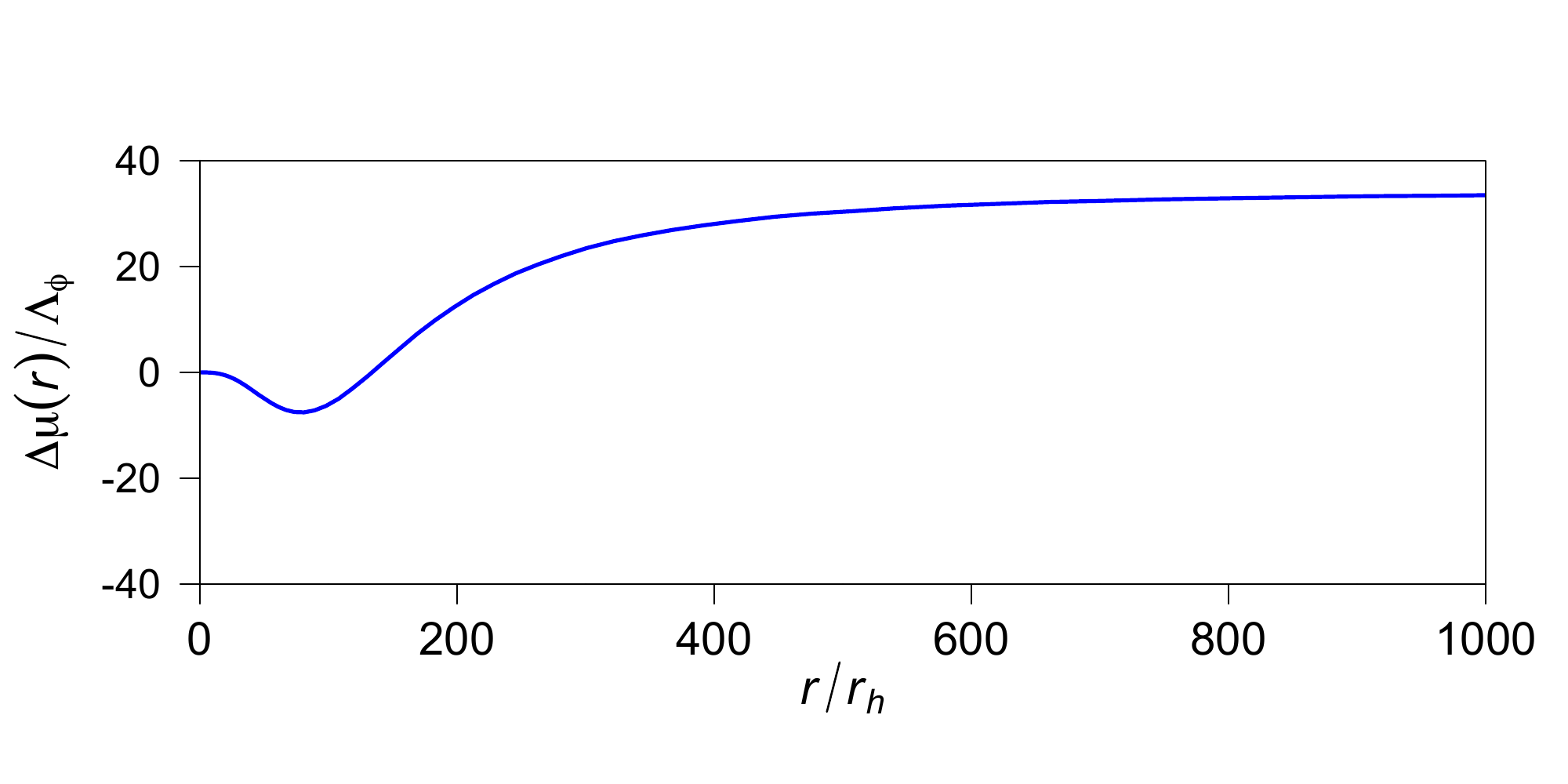}
\caption{Profiles for the bubble and the mass term $\mu(r)$ outside 
the horizon $r_h$ with $M_5=10^{15}$GeV, $\Lambda_{\phi}=10^{12}$GeV 
and $r_h=20000/M_p$.}
\label{fig:field}
\end{figure}

Figure \ref{fig:field} gives profiles for the bubble centered on the black 
hole after tunnelling and for the mass term $\mu(r)$ beyond the horizon 
radius $r_h$. The solutions are shown for an instanton with action $B=4.3$. 
The field is in the true vacuum at the horizon and approaches the false 
vacuum as $r\to\infty$ with a characteristic thick wall profile. 
The bubble radius greatly exceeds the horizon of the black hole.

The change in the mass term is given by $\Delta\mu(r)=\mu(r)-\mu(r_h)$. 
Near the horizon, $\Delta \mu(r)$ is negative due to the negative potential 
$V$ in equation \ref{eq2}. 
$\mu(r)$ becomes positive at large $r$ where there is a positive contribution 
from the kinetic term and hence $\Delta M$ is positive.

\subsection{Branching ratios}

The calculation of the vacuum decay rate assumes a stationary background
which only makes sense when the decay rate exceeds the Hawking evaporation rate.
The brane black hole can radiate in the brane or into the extra dimension, but
if we consider a scenario as close as possible to the standard model then most
of the radiation will be in the form of quarks and leptons radiated into the brane,
simply because these are the most numerous particles. (For a review of Hawking 
evaporation rates in higher dimensions see \cite{Kanti:2014vsa}.)

Black hole radiation is similar to the radiation from a black body with the same area
as the black hole horizon and at the Hawking temperature,
but with additional `grey body' factors representing the effects of back-scattering of
the radiation from the space-time curvature around the black hole.
Following \cite{Kanti:2014vsa}, we can express the energy loss rate due to evaporation 
as $\dot E$, where on dimensional grounds (since $r_h$ is the only relevant dimensionful
parameter)
\begin{equation}
|\dot E|=\gamma \,r_h^{-2},
\end{equation}
for some constant $\gamma$. The Hawking decay rate of the black hole $\Gamma_H$, using
(\ref{mtorh}) to eliminate the radius, is
\begin{equation}
\Gamma_H=\frac{|\dot E|}{M_S}=\frac{4\pi\gamma M_5^3}{M_S^2}
\end{equation}
The vacuum decay rate is given by
\begin{equation}
\Gamma_D=Ae^{-B}.
\end{equation}
The pre-factor $A$ contains a factor $(B/2\pi)^{1/2}$ from a zero mode 
and a vacuum polarisation term from the other modes, whose characteristic 
length scale is the bubble radius $r_b$. We estimate
\begin{equation}
\Gamma_D\approx\left(\frac{B}{2\pi}\right)^{1/2}\frac{1}{r_b}e^{-B}.
\end{equation}
The branching ratio of the two is
\begin{equation}
{\Gamma_D\over\Gamma_H}\approx{1\over\gamma}\left(\frac{B}{2\pi}\right)^{1/2}
\left(\frac{M_S}{M_5}\right)^{3/2}\left(\frac{r_h}{r_b}\right)e^{-B}
\end{equation}
Vacuum decay is important when this ratio is larger than one.

In the case of small $r_h/\ell$, the five-dimensional black hole has a temperature
\begin{equation}
T\approx \frac{1}{2\pi r_h},
\end{equation}
which is double the temperature of a black hole solely in four dimensions. 
We would therefore expect to have energy flux on the brane roughly
$\propto T^4\sim 16$ times the flux solely in four dimensions. Numerical
results actually give a factor of 14.2 for fermion fields, which give the 
largest contribution to the decay \cite{Harris:2003eg}. The energy
loss due to a fermion in four dimensions contributes a factor of 
$7.88\times 10^{-4}$ for each degree of freedom to $\gamma$, 
giving a total for $90$ standard model fermion degrees of freedom of
\begin{equation}
\gamma\approx 14.2\times 90\times 7.88\times 10^{-4}=0.10.
\end{equation}
\begin{figure}[htb]
\centering
\includegraphics[width=0.7\textwidth]{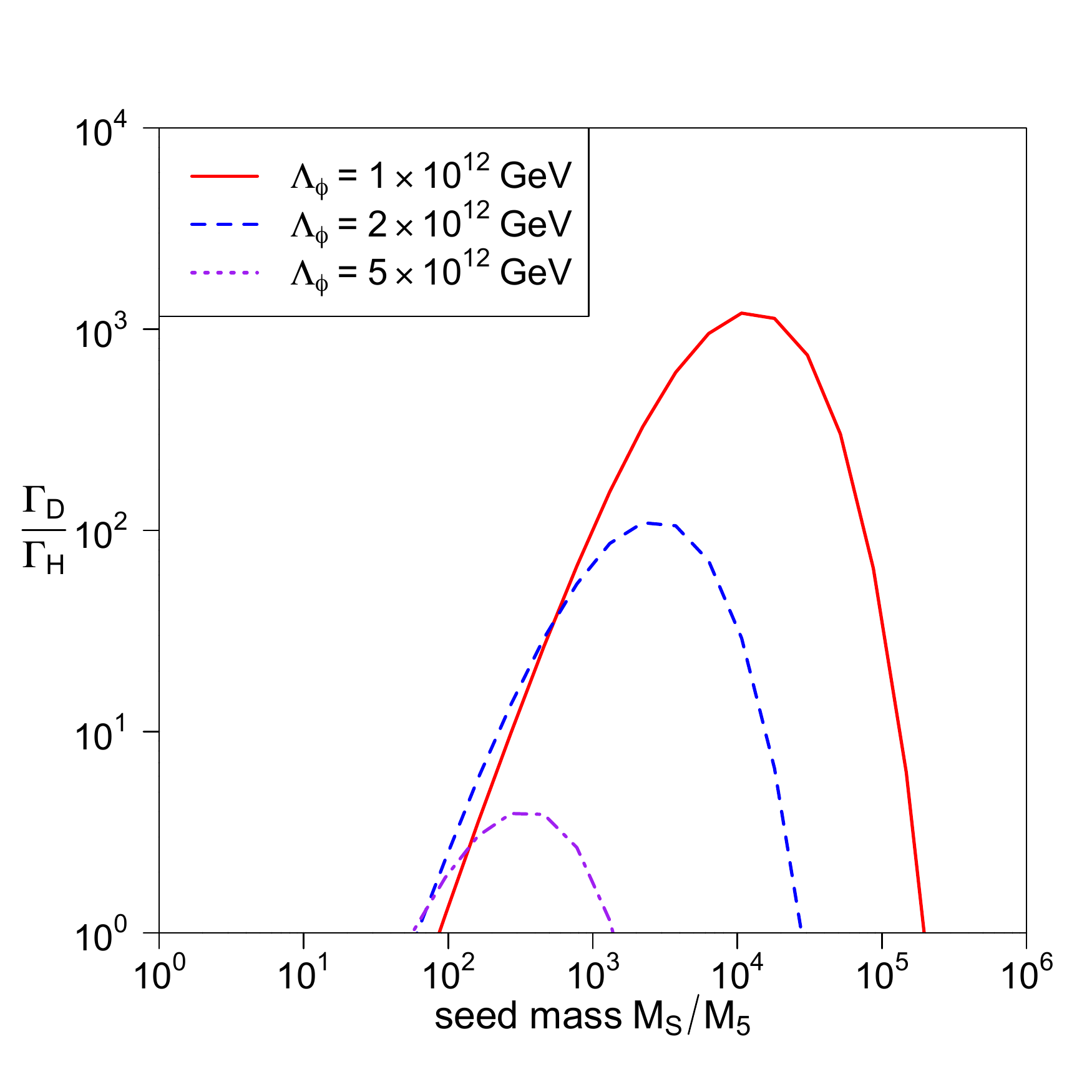}
\caption{
The branching ratio of the false vacuum nucleation rate to 
the Hawking evaporation rate as a function 
of the seed mass for a selection of Higgs models with $M_5=10^{15}{\rm GeV}$.}
\label{fig:ratio}
\end{figure}

The branching ratio is plotted in figure \ref{fig:ratio} for $M_5=10^{15}{\rm GeV}$ and
Higgs instability scale around $10^{12}\,{\rm GeV}$ (corresponding to a top quark
mass of $172\,{\rm GeV}$). Note that the decay rates in this parameter range are
larger than $M_5^3/M_S^2$, i.e.\ they are extremely fast. The figure shows an example
where black holes with masses between $10^{17}\,{\rm GeV}$ and $10^{20}\,{\rm GeV}$, 
or $10^{-7}\,{\rm g}$ to $10^{-4}\,{\rm g}$, would seed rapid  Higgs vacuum decay.

\section{Conclusions}
\label{sec:disc}

In this paper we have explored the impact of large extra dimensions
on black hole seeded vacuum decay. We used the Randall-Sundrum
set-up as a concrete example for warped extra dimensions, and 
numerically computed the Higgs profile on the brane for vacuum decay
assuming a \emph{tidal Ansatz} for the Weyl tensor on the brane. 
Although the solution for a brane black hole is not known analytically, we 
were nonetheless able to construct an argument that the action for tunnelling
would still be the difference in areas of the black hole horizons. In order
to estimate these areas, we focussed on small brane black holes (expected
to be the most relevant for vacuum decay), and used qualitative features of
the numerical solutions to argue the black hole area would be very well 
approximated by the hyperspherical result $2\pi^2r_h^3$. We then used the
tidal model for a brane black hole (in keeping with the tidal Ansatz for the 
Weyl tensor), expanded for small masses, to relate the 4D brane mass of the
black hole, the $1/r$ fall-off of the Newtonian potential, to the horizon radius.
This then allowed us to compute the amplitude for tunnelling.

Since a black hole can also radiate, we then have to consider whether
the evaporation rate is so fast that the tunnelling amplitude is irrelevant,
or whether the tunnelling probability becomes so high for small black holes
(as was the case for purely four dimensional black holes \cite{BGM3}) that
the black hole always initiates decay. We therefore estimated the nett
evaporation rate by taking the integrated flux from \cite{Harris:2003eg},
which is dominated by the fermion radiation, and summing up the effect
from the standard model particles. The branching ratio plot of figure \ref{fig:ratio}
demonstrates that, just as in 4D, small black holes in higher dimensions
are overwhelmingly likely to initiate vacuum decay once they have 
radiated away sufficient mass to enter this danger range. As with pure 4D,
any small black hole, formed either in the early universe, or in a high energy
cosmic ray collision, will radiate, lose mass, then become sufficiently light
that it seeds decay with a rate of order $10^{3-5} T_5$.\footnote{Here,
$T_5 = (c^3/8\pi G_5 \hbar)^{1/3}$ is the 5D Planck time.} What is interesting
here is that what we mean by \emph{small} is now very different to the
pure 4D case.

With large extra dimensional scenarios, we generate a high 4D Planck 
scale geometrically, having a renormalization of the Newton constant
coming from the `volume' of the internal dimensions. Thus, in 4D, where
the typical black hole seeding vacuum decay for the Higgs was
in the range $10^5-10^9 M_p \simeq 1$g$-10$ tonnes, these black
holes could only be primordial in origin, having far too high a mass to be 
produced in a particle collision. Here however, our Planck mass can be much 
lower, so $10^5 M_5$ can potentially be sufficiently low that the black hole
could be produced in cosmic ray collision. For example,
the highest energy cosmic ray collisions
\cite{Linsley:1963km,Nagano:2000ve,ThePierreAuger:2015rha} observed
have an energy in excess of $10^{11}$GeV. 
Hut and Rees \cite{Hut:1983xa} have shown that there are
at least $10^5$ collisions with centre of mass energy exceeding $10^{11}$ GeV in our
past light cone. Thus, provided the higher dimensional
Planck scale were below $M_5\lesssim 10^9$GeV, black holes could
be formed in a cosmic ray collision that would be sufficiently light to
catalyse vacuum decay.

In the context of the Higgs field, the standard model potential is only 
valid at best for energy scales below the scale of new physics, $M_5$,
therefore the instability scale should satisfy $\Lambda_\phi<M_5$.
The lowest possible value for the instability scale consistent with experimental 
limits on the top quark mass is around $10^8\,{\rm GeV}$, thus we cannot use
our standard model Higgs decay results unless $M_5\gg10^8\,{\rm GeV}$,
well outside the range probed by the LHC. 

As an example, consider an instability scale $\Lambda_\phi\sim 10^{8}\,{\rm GeV}$,
and Planck scale $M_5\sim10^{9}\,{\rm GeV}$, then black holes of mass 
$M_S\sim10^{11}\,{\rm GeV}$ could cause Higgs vacuum decay. 
These values are below those for which we were able
to obtain numerical results, but we can make a rough approximation by
taking the exponent for vacuum decay $B$ from \eqref{Bapprox}, and 
the mass of the instanton $\delta M\sim \Lambda_\phi$. For these
values we estimate $B=O(1)$ and rapid Higgs decay would take place.

\begin{figure}[htb]
\centering
\includegraphics[width=0.7\textwidth]{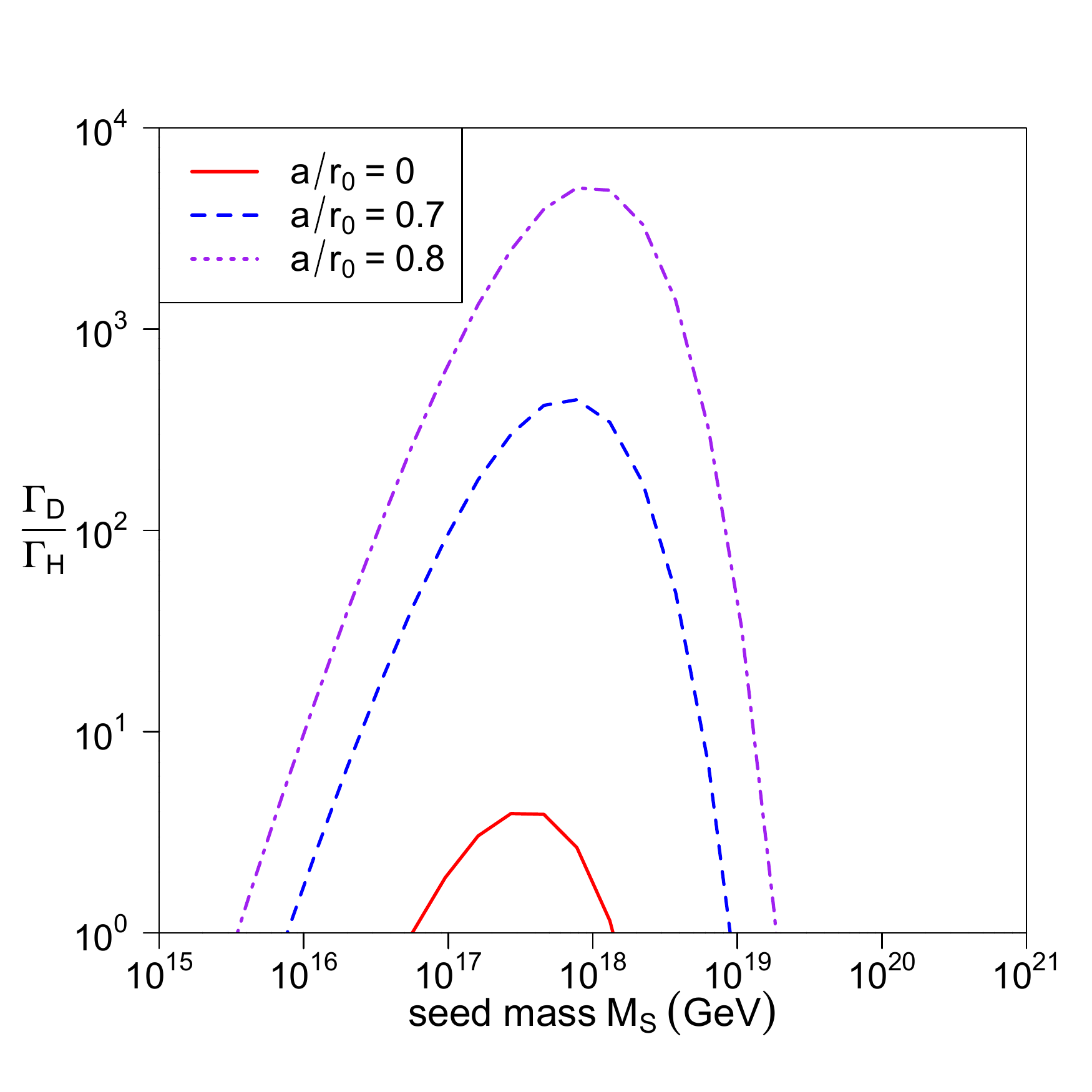}
\caption{
The branching ratio of the false vacuum nucleation rate to 
the Hawking evaporation rate as a function of the seed mass 
for a selection of Higgs models with $M_5=10^{15}{\rm GeV}$,
and $\Lambda_\phi=5\times 10^{12}\,{\rm GeV}$}
\label{fig:rotating}
\end{figure}

Black holes produced by high energy collisions would be likely to 
be rotating. Rotating tidal black hole solutions \cite{Aliev:2005bi} 
can be used as the basis for these black hole seeds. 
The bubble solutions about these rotating holes will become
distorted, however the profile of the bubble solution (fig.\ \ref{fig:field})
indicates that much of the variation of the bubble fields occurs at large
radii compared to the horizon size of the black hole. This
suggests that the distortion will be localised in the 
small part of the bubble near the black hole,
leaving the effective mass $\delta M$ in the field configuration 
relatively unaffected.
In this case, we can use our earlier result (\ref{Bapprox}) but 
replacing the horizon area with the area ${\cal A}_{MP}$ of a rotating 
Myers-Perry black hole in flat space \cite{Myers:1986un} when $r_h\ll \ell$,
\be
B\approx \frac{{\cal A}_{MP}}{4G_5}\frac{3\delta M}{2M_S}.
\ee
The area depends on two rotation parameters $a_1$ and $a_2$, 
but for a rotation axis aligned to the brane we can take $a_2=0$. 
In this case
\be
{\cal A}_{MP}=2\pi^2r_0^3\left(1-\frac{a^2}{r_0^2}\right)^{1/2},
\ee
where $r_0$ is the horizon radius of the non-rotating black hole solution,
\be
r_0^2=\frac{8G_5M_S}{3\pi}.
\ee
The area is smaller than the non-rotating case.
Furthermore, the Hawking temperature is reduced, since
\be
T_H=T_0\left(1- \frac{a^2}{r_0^2}\right)^{1/2}
\ee
The numerical results for vacuum decay are shown in figure \ref{fig:rotating}.
The vacuum decay rate $Ae^{-B}$ with rotating seeds is larger 
than than with non-rotating seeds due to the reduced area.

While this is a rather rough argument, the basic intuition that the 
branching ratio will be enhanced both by the larger decay rate and 
the reduced Hawking evaporation rate is likely to be correct. In other
words, if the existence of large extra dimensions does not destroy the 
vacuum metastability of the standard model Higgs, then ultra high energy
particle collisions risk producing black hole seeds that will catalyse the
decay of the vacuum.

\acknowledgments

We are grateful for the hospitality of the Perimeter Institute, where part
of this research was undertaken. This work was supported in part by the 
Leverhulme grant \emph{Challenging the Standard Model with Black Holes}
and in part by STFC consolidated grant ST/P000371/1.  
LC acknowledges financial support from CONACyT,
RG is supported in part by the Perimeter Institute for Theoretical Physics,
and KM is supported by an STFC studentship.
Research at Perimeter Institute is supported by the Government of
Canada through the Department of Innovation, Science and Economic 
Development Canada and by the Province of Ontario through the
Ministry of Research, Innovation and Science.


\appendix
\section{Canonical decomposition}
\label{Appaction}

In this appendix we review and extend the ideas given in \cite{Hawking:1995fd}
that provide a canonical decomposition of a manifold (in our case a
Euclidean one) by a foliation of hypersurfaces $\Sigma_\tau$ to recast 
the gravitational action in its Hamiltonian version. 

The gravitational equations on a manifold $\mathcal{M}$ with boundary 
$\partial \mathcal{M}$ are obtained by the extremisation of the usual 
Einstein-Hilbert action plus a Gibbons-Hawking surface term:
\begin{equation}
I= - \frac{1}{16\pi G_5} \int_{\mathcal{M}}\left(R_5 - 2\Lambda_5\right)  
\sqrt{g} + \int_{\mathcal{M}} \mathcal{L}_m(g,\phi)  \sqrt{g} 
+ \frac{1}{8\pi G_5} \int_{\partial\mathcal{M} }\sqrt{h}  K,
\label{TBHaction}
\end{equation}
here $\mathcal{L}_m$ is the matter Lagrangian, $h_{ab} = g_{ab}
-n_a n_b$ is the induced metric and 
$K=g^{ab}K_{ab}=g^{ab}h_a{}^c h_b{}^d \nabla_c n_d$ 
is the trace of the extrinsic curvature of the boundary 
$\partial \mathcal{M}$ with normal vector $n_a$ pointing \emph{in}
to ${\cal M}$.

\begin{figure}[htb]
\centering
\includegraphics[width=0.4\textwidth]{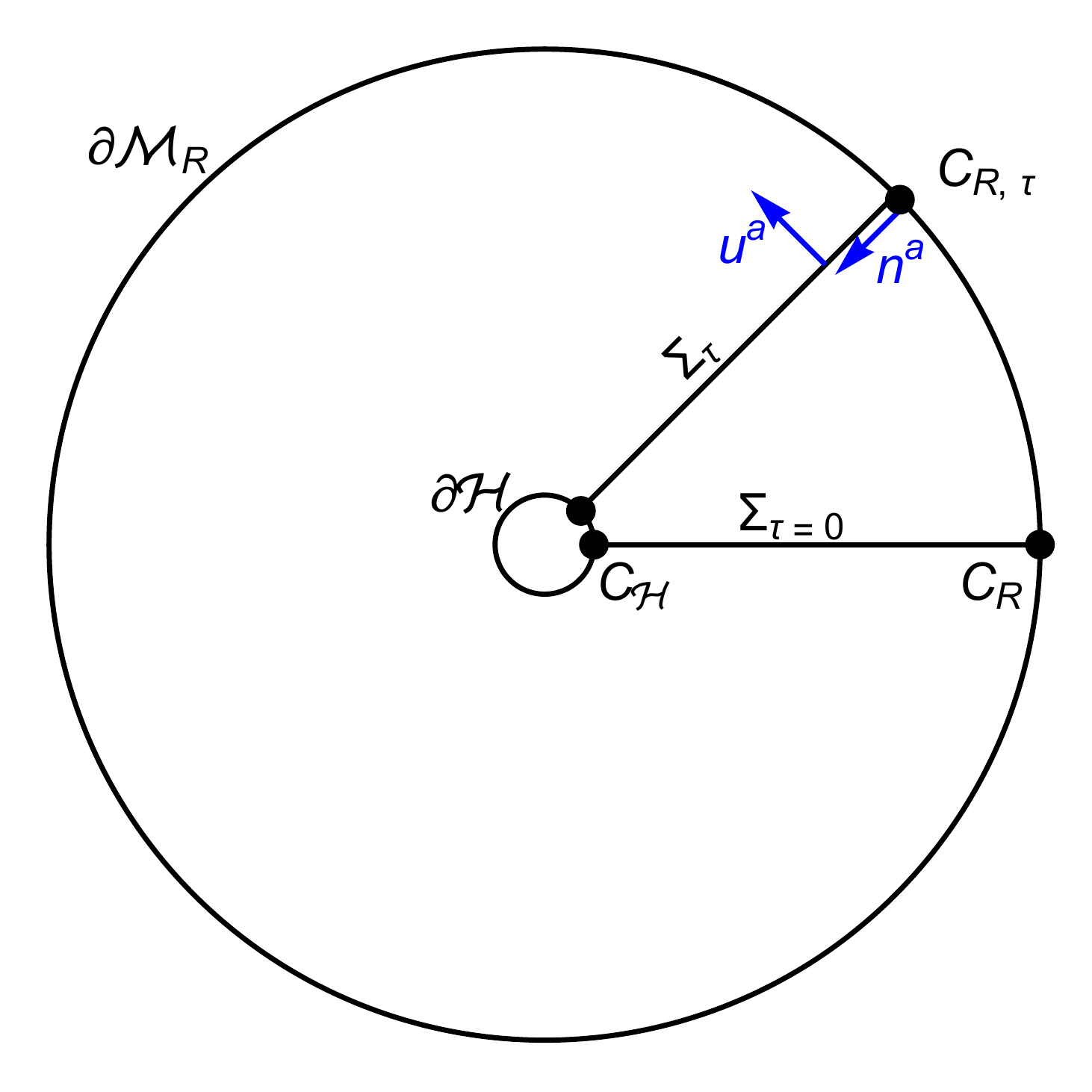}
\caption{
An illustration of the foliation of the Euclidean $\{\tau,r\}$ section
of the brane black hole. The normals $u^a$ and $n^a$ of, respectively, 
the foliation $\Sigma_\tau$ and manifold boundaries are shown, together
with the codimension two surfaces $C_{R,\tau}$ that are regarded as
a codimension one submanifold of the $\Sigma_\tau$ surfaces.}
\label{fig:canon}
\end{figure}

To simplify this action we make a foliation of the spacetime 
$\mathcal{M}$ by codimension one time-slices 
$\Sigma_\tau$, labelled by a periodic Euclidean time function 
$\tau$ which runs from $\tau=0$ to $\tau=\beta$. 
The induced metric on the time-slices is written as
\begin{equation}
\mathfrak{h}_{ab}=g_{ab} - u_a u_b,
\end{equation}
where $u^a$ is a unit normal vector to the slice $\Sigma_\tau$.
In general,  $\partial/\partial\tau$ and $u^a$ will not be 
aligned, but we can decompose $\partial/\partial\tau$ into components 
along the normal and tangential directions, 
\begin{equation}
\left ( \frac{\partial \;}{\partial\tau} \right) ^a = N u^a + N^a
\label{TBHtau}
\end{equation}
The \textit{lapse function}, $N$, measures the rate of flow of 
proper time with respect to the coordinate 
time $\tau$ as one moves through the family of hypersurfaces.
We construct the time-slices $\Sigma_\tau$  to meet the boundary  
$\partial \mathcal{M}$ orthogonally
for convenience. In the case of the region outside the horizon  
for $I_R^{\rm ext}$ (\ref{ir}), the boundary $\partial \mathcal{M}$ is 
composed of two surfaces of constant radius, $\Sigma_{\cal H}$ near
the horizon, and $\Sigma_R$ at large radius.

We use the Gauss identity to relate the Riemann tensor of $g_{ab}$ 
in five dimensions to the Riemann tensor of $\mathfrak{h}_{ab}$ in four, 
and the extrinsic curvatures of the constant time slices 
${\cal K}_{ab}=\mathfrak{h}^c{}_a \mathfrak{h}^d{}_b \nabla_c u_d$, as
\begin{equation}
R_4{}^a_{~bcd}= 
\mathfrak{h}^a{}_{a'} \mathfrak{h}_b{}^{b'} \mathfrak{h}_c{}^{c'} 
\mathfrak{h}_d{}^{d'} R_5{}^{a'}_{~b'c'd'}    
+  {\cal K}^a{}_c {\cal K}_{db} - {\cal K}^a{}_d K_{cb} . 
\label{TBHRiemdec}
\end{equation}
Notice this ${\cal K}$ is distinct from the extrinsic curvature  
of $\Sigma_R$ in \eqref{TBHaction}.
Contracting \eqref{TBHRiemdec} gives 
\begin{equation}
R_5 =R_4 + 2  R_{5ab} u^a u^b - ({\cal K}^2 -{\cal K}^{ab}{\cal K}_{ab}),
\end{equation}
and we obtain a relation between the second term of this expression 
and the extrinsic curvature by commuting covariant derivatives of the 
normal vector
\begin{equation}
R_{5ab} u^a u^b = 2 u^b \nabla_{[c}\nabla_{b]}u^c
= {\cal K}^2 - {\cal K}^{ab} {\cal K}_{ab} - \nabla_{a}(u^a \nabla_{c}u^c) 
+ \nabla_{c}(u^a \nabla_{a}u^c).
\end{equation}
Combining these two expressions leads to the identity,
\begin{equation}
R_5 =R_4 - ({\cal K}^{ab} {\cal K}_{ab} - {\cal K}^2) - 
2 \left[  \nabla_{a}(u^a \nabla_{c}u^c) - \nabla_{c}(u^a \nabla_{a}u^c) \right],
\label{TBHRiccidec}
\end{equation}
which forms the basis of all canonical decompositions of the Einstein-Hilbert action.

When substituted in \eqref{TBHaction}, the last two terms of 
\eqref{TBHRiccidec} are reduced to boundary contributions on 
$\partial \mathcal{M}$. The first of these vanishes due to orthogonality
of $\partial \mathcal{M}_R$ and $\Sigma_\tau$. 
The second combines with $\int_{\partial {\cal M}} K$ 
from the original action, and gives on $\partial{\cal M}_R$
(with a similar expression for $\partial{\cal H}$)
\bea
\frac{1}{8\pi G_5} \int_{\partial{\cal M}_R} d^4x \sqrt{h}
\left( \nabla_a n^a + n_b u^a \nabla_a u^b \right) &=&
\frac{1}{8\pi G_5} \int_{\partial{\cal M}_R} d^4x \sqrt{h} 
(g^{ab} - u^a u^b) \nabla_a n_b \nonumber \\
&=& \frac{1}{8\pi G_5} \int_{\partial{\cal M}_R} d^4x \sqrt{h}~ 
\mathfrak{h}^{ab} \nabla_a n_b ,
\label{TBHextcurv}
\eea
but this four dimensional integral can be viewed as an integral over 
$\tau$ of a three dimensional integrand that is precisely the three
dimensional extrinsic curvature ${}^3K$ of a family of surfaces 
$C_R(\tau) = \partial{\cal M}_R \cap \Sigma_\tau $ living in the 
boundary $\partial{\cal M}_R$.
A similar term is obtained for the $\partial{\cal H}$ surface near the 
horizon however, for the black hole metrics, it turns out that $^3 K\to 0$ 
as $r\to r_h$, and so this term does not contribute to the action.

Noticing that $\sqrt{g}=N\sqrt{\mathfrak{h}}$, and introducing a metric 
${}^3\mathfrak{h}$ on $C_R$,
we can divide the spacetime integral into space and time,
to express the action \eqref{TBHaction} as
\be
\beal
I= - \int N d\tau \left\{ \frac{1}{16\pi G_5} \int_{\Sigma_\tau} \sqrt{\mathfrak{h}}
\left[ R_4 - ({\cal K}^{ab} {\cal K}_{ab} - {\cal K}^2)-2\Lambda_5 
- 16\pi G_5 \mathcal{L}_m \right]  \right.\\ 
\left. - \frac{1}{8\pi G_5} \int_{C_R} \sqrt{{}^3\mathfrak{h}} ~ ^3 K 
- \frac{1}{8\pi G_5} \int_{C_{\cal H}} \sqrt{{}^3\mathfrak{h}} ~ ^3 K \right\}.
\hskip 2cm
\eeal
\label{TBHactiondec}
\ee
Furthermore, we can see how the extrinsic curvature is related to the 
Lie derivative of the intrinsic metric with respect to $\tau$ via 
\eqref{TBHtau}:
\begin{eqnarray} 
{\cal K}_{ab}=\frac{1}{2} \mathsterling_u \mathfrak{h}_{ab}
=\frac{1}{2N} \left( \mathsterling_\tau \mathfrak{h}_{ab} -
\mathsterling_N \mathfrak{h}_{ab} \right) = 
\frac{1}{2N} \left( \dot{\mathfrak{h}}_{ab} - 2 D_{(a}N_{b)} \right),
\label{TBHextreln}
\end{eqnarray} 
where $\dot{\mathfrak{h}}_{ab} = \mathfrak{h}^c_a \mathfrak{h}_b^d 
\mathsterling_\tau \mathfrak{h}_{cd}$ 
and $D_a$ is the derivative associated with $\mathfrak{h}_{ab}$. 

To obtain the Hamiltonian form of $I$ we define the canonical 
momentum $\pi^{ab}$ conjugate to the intrinsic metric as
\begin{equation}
\pi^{ab} \equiv \frac{\delta I }{\delta \dot{\mathfrak{h}}_{ab}}= 
\sqrt{\mathfrak{h}} ({\cal K}^{ab} - {\cal K} \mathfrak{h}^{ab}),
\label{TBHcanmom}
\end{equation}
This allows us to recast \eqref{TBHactiondec} in terms of the canonical momentum
\begin{gather}
I= - \int_0^\beta N d\tau \left\{ 
\frac{1}{16\pi G_5}\int_{\Sigma_\tau} \sqrt{\mathfrak{h}} 
\left[ R_4 -  \frac{1}{\mathfrak{h}}\left( \pi^{ab}\pi_{ab} - 
\frac{1}{3} \pi^2 \right)-2\Lambda_5 - 16\pi G_5 \mathcal{L}_m \right]
\nonumber\right.\\ 
\left. - \frac{1}{8\pi G_5} \int_{C_R} \sqrt{\mathfrak{{}^3h}} ~ ^3 K 
- \frac{1}{8\pi G_5} \int_{C_{\cal H}} \sqrt{\mathfrak{{}^3h}} ~ ^3 K  \right\}.
\end{gather}

Now we are ready to perform a Legendre transformation of the 
Lagrangian and using \eqref{TBHextreln} and
\eqref{TBHcanmom} to obtain the Hamiltonian formulation. 
\be
\beal
I = \frac{1}{8\pi G_5}\int_0^\beta d\tau \left\{ 
\frac12\int_{\Sigma_\tau} \sqrt{\mathfrak{h}} 
\left( \pi^{ab}\dot{\mathfrak{h}}_{ab} - N \mathcal{H} - N^a \mathcal{H}_a  \right)
\right. \hskip 3cm \\ 
\left. +  \int_{C_R} \sqrt{{}^3\mathfrak{h}} 
( N ~ ^3 K  + N^a \pi_{ab} n^b) 
+ \int_{C_{\cal H}} \sqrt{{}^3\mathfrak{h}} 
( N ~ ^3 K  + N^a \pi_{ab} n^b) \right\},
\eeal
\ee
with the \textit{Hamiltonian constraint} function $\mathcal{H}$ and 
the \textit{momentum constraint} function $\mathcal{H}^a$ given by
\be
\beal
\mathcal{H}^a &= -2 D_b \left(\frac{1}{\sqrt{\mathfrak{h}}} \pi^{ab} \right) \\
\mathcal{H} &=  R_4 - 2\Lambda_5 + \frac{1}{\mathfrak{h}} 
\left( \pi^{ab}\pi_{ab} - \frac{1}{3} \pi^2 \right)  - 16\pi G_5 \mathcal{L}_m.
\eeal
\label{TBHham}
\ee

Finally, for a static spacetime we have  $\dot{\mathfrak{h}}_{ab}=0$ 
and in the non-rotating case $N^a=0$. The metric is a solution
to the field equations, so that in particular we have the constraint 
equations ${\cal H}={\cal H}^a=0$.
The only non-vanishing part of the action are the two boundary terms $^3K$,
\be
I = \frac{1}{8\pi G_5} \int_0^\beta d\tau 
\left( \int_{C_R}  {}^3 K\, \sqrt{h} + \int_{C_{\cal H}}  {}^3 K\, \sqrt{h} \right).
\ee
For our black hole solutions, this diverges in the limit $R\to 0$. 
However, the matter contributions to the black hole instanton solutions 
die off exponentially at large radii, so that the boundary terms cancel when 
we calculate the difference in actions between the instanton solutions and 
the false vacuum solutions with the same mass and periodicity $\beta$.

\section{Brane equations for the instanton bubble}
\label{app:braneq}

Following the work done in \cite{Shiromizu:1999wj, Dadhich:2000am} 
we briefly review the derivation of the equations (\ref{eq1}-\ref{eq3}), 
which describe the dynamics of the bubble-brane system 
analysed on Section \ref{sec:bubble}.

The Einstein equations for a five dimensional RS braneworld 
can be written as
\begin{equation}
^{(5)}G_{ab} = - \Lambda_5 g_{ab}  + 8\pi G_5  \delta(z) ( -\sigma h_{ab} + T_{ab}) ,
\end{equation}
where $z$ is a coordinate defined by taking the proper distance from
the brane into the bulk, $G_5 = G_N \ell$ and the cosmological constant of the 
bulk $\Lambda_5=-6/\ell^2$ is given in terms of the AdS$_5$ radius $\ell$. 
Notice that we use latin indices for the bulk spacetime whereas 
greek indices will be reserved for objects living on the brane. 
The brane is located at $z=0$ and has an induced metric $h_{ab}$, defined by
\begin{equation}
h_{ab}= g_{ab} - n_a n_b
\end{equation}
where $n^a$ is a unit vector in the $z-$direction. The energy momentum 
tensor of the brane carries the effect of the tension $\sigma$ and has a 
contribution $T_{ab}$, coming from the fields living in the brane.

The Israel junction conditions for the brane allow us to 
write down a set of four dimensional Einstein equations (see \cite{Shiromizu:1999wj}),
\begin{equation}
G_{\mu\nu} = 8 \pi G_N \tilde{T}_{\mu\nu} - {\cal E}_{\mu\nu}
- \Lambda_{\text{eff}} h_{\mu\nu},
\label{effeins}
\end{equation}
where $\Lambda_{\text{eff}}$ is an effective four dimensional cosmological 
constant on the brane, 
\begin{equation}
\Lambda_{\text{eff}} = -\frac{3}{\ell^2} 
+ \frac{(4\pi G_5\sigma)^2}3 \,,
\end{equation}
and ${\cal E}_{\mu\nu}$ is the projection of the five dimensional 
Weyl tensor onto the brane
\begin{equation}
{\cal E}_{\mu\nu} = ~^{(5)}C^\alpha_{~\beta\rho\sigma} n_\alpha 
n^\rho h_\mu{}^\beta h_\nu{}^\sigma ,
\end{equation}
carrying information about the extra dimensional geometry to the brane.
Due to the properties of the Riemann tensor, ${\cal E}_{\mu\nu}$ 
is traceless and divergence free.
In the critical RS brane that will be our false vacuum, the tension of the brane
is tuned so as to set $\Lambda_{\text{eff}}$ to zero, i.e.\ 
\begin{equation}
\sigma=\frac{3}{4\pi G_5 \ell} \; .
\end{equation} 
Finally, the effective energy momentum tensor, $\tilde{T}_{\mu\nu}
= T_{\mu\nu} + \pi_{\mu\nu} $ consists of the standard energy momentum
tensor, together with second order terms 
\begin{equation}
\pi_{\mu\nu} = \frac{1}{ \sigma} \left( -\frac32 T_{\mu\alpha}T^\alpha_\nu + 
\frac12 T T_{\mu\nu} + \frac34 h_{\mu\nu} T_{\alpha\beta} 
T^{\alpha\beta} - \frac14 h_{\mu\nu} T^2 \right).
\end{equation}  

As discussed in section \ref{sec:bubble}, we consider 
static, spherically symmetric solutions on the brane, with
metric \eqref{genbranemet}, and make the tidal Ansatz for
the Weyl tensor:
\begin{equation}
{\cal E}_{\mu\nu} dx^\mu dx^\nu = {\cal U}(r) 
\left( f e^{2\delta} d\tau^2 + f^{-1} dr^2 - r^2 d\Omega^2_{I\!I} \right)
\end{equation}
where the conservation equation gives
\begin{equation}
{\cal U}(r) = - \frac{r_Q^2}{r^4}.\label{rhoE}
\end{equation} 

The metric functions $f(r)$ and $\delta(r)$ are determined by the 
effective Einstein equations \eqref{effeins}. Following \cite{BGM3},
we define a ``mass function'' $\mu(r)$ by
\begin{equation}
f=1 - \frac{2 G_N \mu(r)}{r} - \frac{r_Q^2}{r^2},
\label{appdeff}
\end{equation}
where we have explicitly factored out the tidal term $r_Q^2/r^2$.
The relevant components of the Einstein tensor are
\be
G^t{}_t = - \frac{2G_N \mu'}{r^2} +\frac{r_Q^2}{r^4}\;\;,\;\;\;
G^r{}_r-G^t{}_t = \frac{2f}{ r}\delta'
\ee
For the instanton scalar profile with potential $V(\phi)$, 
the energy-momentum tensor for the scalar field is
\begin{equation}
T_{\mu\nu} = \phi'^2 \delta_\mu^r \delta^r_\nu 
- h_{\mu\nu} \left(\frac12 f \phi'^2 + V \right),
\end{equation}
thus inputting the form of $f$, we see that the tidal contribution is 
cancelled by the tidal tensor, and we finally obtain the equations of motion
(\ref{eq1}-\ref{eq3}) used in the numerical integration:
\be
\beal
0&= f \phi''  + \frac{2}{r} f \phi' +
\delta' f \phi' + f' \phi' - \frac{\partial{V} }{ {\partial \phi}} \\
\mu'(r) &= 4\pi r^2 \left[ \frac12 f \phi'^2 +V 
- \frac{2 \pi G_N}{3} \ell^2 ( \frac12 f \phi'^2 - V )
(\frac32 f \phi'^2 + V)  \right] ,\\
\delta ' &= 4 \pi G_N r \phi'^2 \left[ 1- \frac{4 \pi G_N}{3} \ell^2 ( \frac12 f \phi'^2 - V )
\right].
\eeal
\ee

\providecommand{\href}[2]{#2}
\begingroup\raggedright\endgroup

\end{document}